\documentclass{aa}  

\usepackage{graphicx}

\usepackage{txfonts}
\usepackage{threeparttable}

\usepackage[colorlinks=true,allcolors=blue]{hyperref}

\usepackage[dvipsnames]{xcolor}

\definecolor{orcidlogocol}{HTML}{A6CE39}
\newcommand{\orcidlink}[1]{\href{https://orcid.org/#1}{\textsuperscript{\includegraphics[width=8pt]{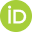}}}}

\begin{document}

\title{Broad Iron Line as a Relativistic Reflection from Warm Corona in AGN}

   \author{P. P. Biswas \orcidlink{0009-0004-6142-9668}
          \inst{1}
          \and
          A. Różańska \orcidlink{0000-0002-5275-4096}
          \inst{1}
          \and
          F. H. Vincent \orcidlink{0000-0002-9214-0830}
          \inst{2}
          \and
          D. Lančová \orcidlink{0000-0003-0826-9787}
          \inst{1,3}
          \and
          P. T. Zycki \orcidlink{0000-0002-7596-4221}
          \inst{1}
          }

   \institute{Nicolaus Copernicus Astronomical Center, Polish Academy of Sciences, Bartycka 18, PL-00-716 Warszawa, Poland\\
              \email{pbiswas@camk.edu.pl}
              \and
              ESIA, Observatoire de Paris, PSL Research University, CNRS, Sorbonne Universités, UPMC Univ. Paris 06, Univ. Paris Diderot, Sorbonne Paris Cité, 5 place Jules Janssen, F-92195 Meudon, France
            \and
             Research Center for Computational Physics and Data Processing, Institute of Physics, Silesian University in Opava, Bezru\v{c}ovo n\'{a}m. 13, 746 01 Opava, Czech Republic
             }

   \date{Received ...; accepted ...}

    \abstract
   {We present that the broad feature usually observed in X-ray spectra at around 6.4~keV can be explained by the ray-traced emission from a two-slab system containing a dissipative, warm corona on the top of an accretion disk in an Active Galactic Nucleus (AGN). Such an accretion flow is externally illuminated by X-ray radiation from a lamp located above a central supermassive black hole (SMBH). Thermal lines from highly ionized iron ions (FeXXV and FeXXVI) caused by { both: an internal heating and}~a reflection from the warm corona, can be integrated into an observed broad line profile due to the close vicinity of the SMBH.}
  {We investigate the dependence of the total broad line profile on the variations in the black hole spin parameter, the viewing angle, lamp height, and dissipation factor. Our results potentially introduce a new method to probe properties of the warm corona using high-resolution spectroscopic measurements with current {\it XRISM} and future {\it NewATHENA} X-ray missions.}
  {We use the photoionization code \texttt{TITAN} to compute local ion population and emission line profiles, and publicly available ray-tracing code \texttt{GYOTO} to include the relativistic effects on the outgoing X-ray spectrum.}
  {In our models, the temperature of the inner atmosphere covering a disk can reach values of $10^7 - 10^8$~K { due to internal warm corona dissipation and external illumination}, which is adequate for generating the highly ionized iron lines.
  These lines can undergo significant gravitational redshift near the black hole, leading to a prominent spectral feature centered around 6.4~keV.} 
   {For all computed models, the relativistic corrections shift highly ionized iron lines to the 6.4~keV region, usually attributed to the fluorescent emission from { the illuminated skin} of an accretion disk. Hence, in the case of the warm corona covering the inner disk regions, the resulting theoretical line profile under the strong gravity is a sum of different iron line transitions, and those originating from highly ionized iron contribute the most to the observed total line profile in AGN.}


   \maketitle
%

\section{Introduction}

 AGN are one of the most prominent X-ray sources. The existence of a reflection component with Fe K$\alpha$ fluorescent emission was first discussed by \citet{1988ApJ...335...57L}. The features were initially reported by \citet{1990Natur.344..132P}, implying a cold accretion disk { with a temperature around} $10^5$ K \citep{1988MNRAS.233..475G} emitting the fluorescent iron line \citep[FeI-FeXVI,][]{1982ApJS...50..263K} { caused by a reflection from neutral material}. With an improved resolution of 
 {\it XMM-Newton} and {\it Chandra} { X-ray telescopes}, a broad iron line was reported with a narrow core \citep[e.g.][]{1995ApJ...453L..81Y,1997ApJ...477..602N,2003ASPC..290...35R,2006AN....327.1032G}, which suggests a relativistic broadened feature originating from the inner accretion disk, while the narrow core being emitted from a distant region \citep[see][]{2000PASP..112.1145F,2006MNRAS.368L..62N,2007MNRAS.382..194N,2022A&A...664A..46A}. The broadening of the fluorescent Fe K$\alpha$ line { generated very close to the supermassive black hole (SMBH)}
 is subjected to relativistic effects like rotational broadening, relativistic boosting, and gravitational redshift, which generates the characteristic double-peaked, { disk-like}, emission line \citep{1991laor}. { Thus, the overall} shape of the line profile depends on the spin of the SMBH and the observer's viewing { angle} \citep{2000PASP..112.1145F}. 

 A commonly accepted scenario is that the observed iron line { is created during the reflection of hard X-rays, originating from an external source, from the inner regions of an accretion disk} \citep{1989fabian,1991george}. { Strong illumination supports the formation of so-called `ionized skin' on the top of accretion disk \citep{1993Ross,2000nayakshin,2000Madej,2001Ballantyne,2002rozanska}.} { Even if an `ionized skin' can pass through different ionization states of iron, the heated top layer was always optically thin and it was a subject to thermal instabilities} \citep[][and references therein]{2000nayakshin,2001Nayakshin,2002rozanska}. { Models of `ionized skin', like \texttt{relxill} \citep[][and references therein]{2014garcia} are commonly used to interpret X-ray reflected spectra of AGN.}

{ Nevertheless, \citet{1999Rozanska} showed that an `ionized skin' is relatively optically thick.} But recent X-ray data show the existence of soft X-ray excess in the majority of AGN \citep[e.g.][]{2016A&A...588A..70B,2020MNRAS.491..532G}. The origin of such excess is still under debate, but recent studies strongly favor the presence of a warm corona, i.e., high-temperature ($k_{\rm B}T \sim 2$ keV, where $k_{\rm B}$ is the Boltzmann constant) structure close to the SMBH. A warm skin on top of the accretion disk 
\citep{2013A&A...549A..73P} is an important part of the so-called two-corona model \citep{2018A&A...611A..59P}: i.e., fully ionized hot corona inner structure as an origin of observed power-law emission, and a warm corona as an origin of soft X-ray excess. 
Final spectrum from such a model consists of soft excess together with power-law, and it was successfully used for interpretation of high-resolution X-ray data of many AGN \citep{2012MNRAS.420.1848D,2018A&A...611A..59P,2024MNRAS.530.1603B,2024A&A...690A.308P}. It was { analytically shown} by \citet{2015A&A...580A..77R} that to produce a warm optically thick layer, covering a cold accretion disk, the inclusion of an internal dissipation in the warm corona is required, in addition to radiation and strong Compton cooling. {This suggests that the emissivity profile would deviate from the prediction of a standard disk atmosphere.} Furthermore, radiation magneto-hydrodynamic simulations of the accretion disk around an SMBH indicated the existence of a dissipative warm/hot slab as an upper zone of the accretion disk atmosphere \citep[e.g.][and references therein]{2019ApJ...885..144J}.

As the temperature in the warm corona is high, iron atoms are likely to be highly ionized. { The warm corona temperatures, of the order of $\sim 2$\,keV}, are suitable for the emission of Fe K$\alpha$ from highly ionized species like FeXXV and FeXXVI \citep{2002A&A...387...76B}. It is worth mentioning that from theoretical point of view, if the K$\alpha$ transition originates from the high-ionization states of iron like FeXVII–FeXXIV, the line photons undergo resonant Auger destruction, significantly reducing their emission intensity \citep{2005AIPC..774...99L}. We also expect the warm corona to be very close to the SMBH, due to which the FeXXV and FeXXVI iron lines are subjected to strong gravitational effects. A broad emission feature, likely from the highly ionized iron ions, was already reported by \citet{2001A&A...365L.134R} and \citet{2001ApJ...559..181P}.  If the highly ionized Fe K$\alpha$ lines are generated very close to the SMBH, then we also expect them to get gravitationally redshifted moving away from their core emission region, broadened due to rotation, and Doppler boosting. Hence, the broad line observed around the 6.4 keV region may be composed of many transitions on highly ionized iron ions.

\begin{figure}[]
\centering
    \includegraphics[width=0.98\linewidth]{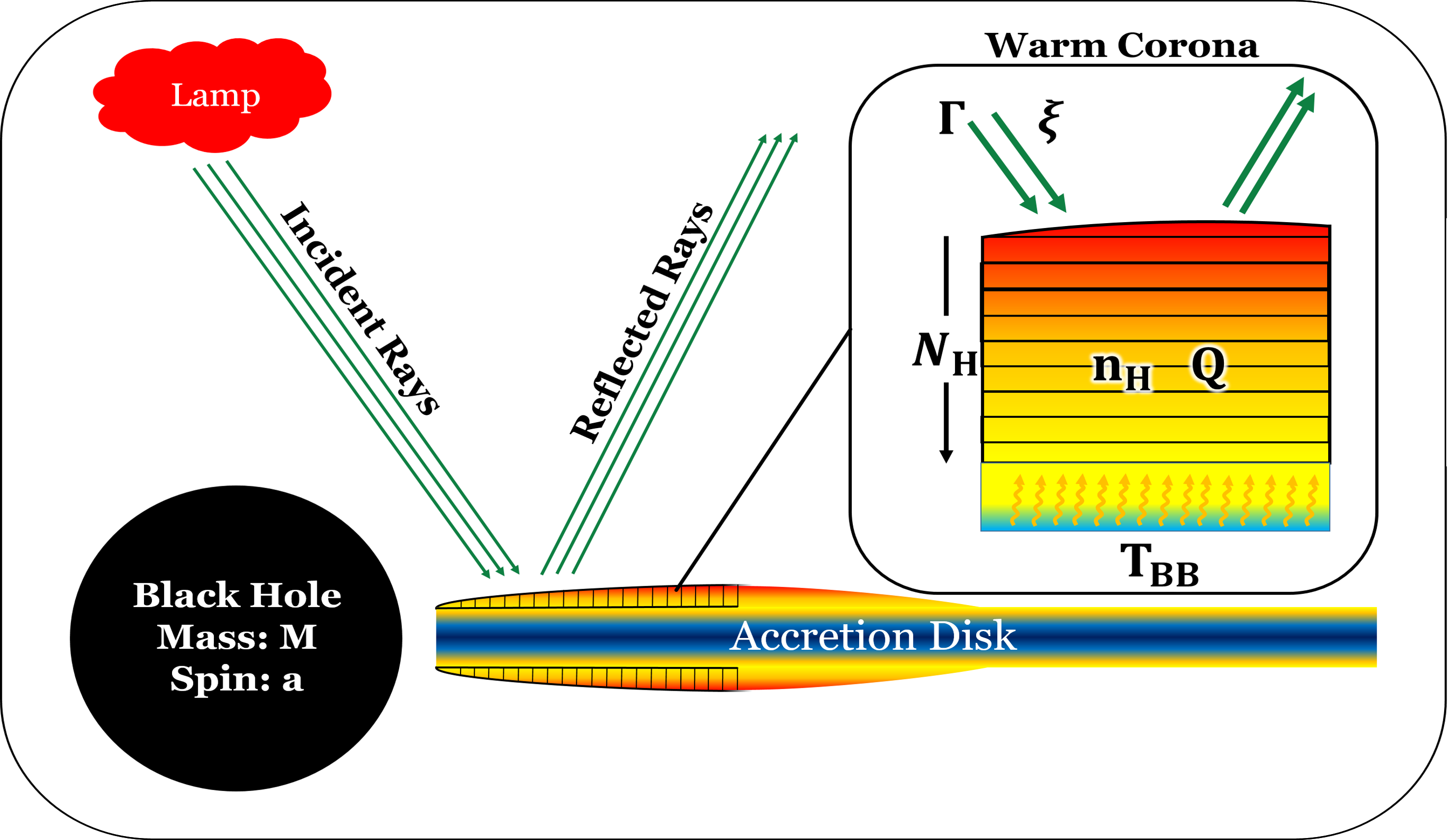}
    \caption{A schematic diagram of our model's setup. A hot corona, as in lamp post geometry, is located at height $h$ above the SMBH of mass $M$ and spin $a$. The warm corona, which is radially and vertically stratified, is positioned on top of an accretion disk. The incident flux coming from the hot corona is reprocessed in the warm corona, which is then detected by the observer. The $r_{\rm in}$ and $r_{\rm out}$ are the integration limits for our model. For a give distance $r$ we compute the vertical structure using \texttt{TITAN} with the input parameters depicted in the warm corona panel; $\Gamma$ - photon index, $\xi$ - ionization parameter, $N_{\rm H}$ - column density which is analogous to the optical depth $\tau$, $n_{\rm H}$ - gas number density, $Q$ - internal heating, and $T_{\rm BB}$ - the temperate of black body radiation from the disk. }
    \label{fig:Schem_dig}
\end{figure}


In this work, we show the effects of strong gravity subjected to the highly ionized Fe K$\alpha$ profiles produced in the warm corona. { To do so}, we consider the two-corona model schematically presented in Fig.~\ref{fig:Schem_dig}.
The radiation from hot corona illuminates a two-zone structure: an accretion disk that is surrounded by a dissipative, Compton-cooled warm corona. 
In addition, a warm atmosphere being on the top of an accretion disk is 
radially stratified. Due to its dissipative nature, the inner regions of the disk atmosphere may reach temperatures characteristic of warm corona at a large optical depth
\citep{2015A&A...580A..77R,2020A&A...634A..85P}. At such high temperatures, the spectrum is dominated by the emission of highly ionized Fe K$\alpha$ ions (FeXXV, FeXXVI) \citep{2020MNRAS.491.3553B}. { Contrary to the `ionized skin' that is only heated by external illumination, warm corona is additionally heated by internal mechanical process, which ensures its high temperature from the outset}. The highly ionized Fe K$\alpha$ lines produced in the warm corona close to the SMBH would be relativistically redshifted, broadened due to fast rotation of the disk, and relativistically boosted, appearing as a broad Fe K$\alpha$ structure. To model highly ionized iron line profiles in the close vicinity of an SMBH, we used a combination of the photoionization code \texttt{TITAN} \citep{2000A&A...357..823D, 2003A&A...407...13D}, and the publicly available ray tracing code \texttt{GYOTO}\footnote{\href{https://gyoto.obspm.fr/index.html}{https://gyoto.obspm.fr/index.html}} \citep{2011CQGra..28v5011V}.
Radiative transfer code \texttt{TITAN} produces angle-dependent intensities with all relevant iron lines, illuminated in the warm corona on top of the cold disk. Those intensities act as the input for \texttt{GYOTO} to compute the final ray-traced spectrum seen by the distant observer. This form of modeling enables us to include relativistic effects in the computed spectrum for different viewing angles and black hole spins.
{ \texttt{GYOTO} code was already used to ray trace the reflected spectra from X-ray binaries \citep{2016A&A...590A.132V}. In addition to the proper treatment of all spectral effects, as many other convolution ray-tracing models for X-ray data analysis, \texttt{GYOTO} also produces images at the assumed energy range, which we present here.}

{ The goal of this paper is to follow the same scheme of computations as in \texttt{reXcor} model of \citet{2024MNRAS.530.1603B}, but with the use of different computational codes. Nevertheless, the global disk solutions for its radial structure remain the same. Since our radiative transfer code \texttt{TITAN} contains a rich set of iron transitions and angle-dependent emission, we can deeply investigate the region of the iron line complex in the reflected spectrum. In this paper, the Compton scattering is included in the total energy balance, and therefore, the gas temperature is determined correctly. But we decided not to include the broadening of the lines due to Compton scattering in order to first study purely relativistic broadening.} The model description on how to combine \texttt{TITAN} and \texttt{GYOTO} computations with their assumptions is discussed in Section \ref{sec:mod_des}. The results are presented in Section \ref{sec:results}. And finally, Section \ref{sec:dic_con} is dedicated to discussion and conclusions.

\section{Model Description} \label{sec:mod_des}

We calculate the observed emission from the inner part of the accretion disk atmosphere in plane-parallel geometry.
A schematic representation of our model is depicted in Fig.~\ref{fig:Schem_dig}. We assume a dissipative warm corona on top of a geometrically thin optically thick accretion disk. We adopt a lamppost geometry, where the disk is illuminated by a source placed at a height $h$ above the black hole, along the spin axis \citep[e.g.][]{1991A&A...247...25M,1996MNRAS.282L..53M,2013MNRAS.430.1694D}. 
The lamp represents a hot corona with emission as a power-law, which illuminates the warm optically thick corona in the direction normal to the surface. Nevertheless, the effect of gravitational light bending of the illuminated radiation is schematically taken into account in Eq.~\ref{eq:FX_r} below. {We note that the lamp approximation holds given that the height of the lamp is not too small, where the light bending is extreme.}

In our calculation of the observable emission, we chose a region very close to the SMBH where the relativistic effects are dominant, and the relatively high temperature of the warm corona can produce highly ionized Fe K$\alpha$ ions. 
For the assumed black hole mass -- $M$, dimensionless spin -- $a$, and disk accretion rate -- $\lambda$, we calculate radially dependent parameters of gas and radiation defined on the top of the atmosphere given in Sec.~\ref{sec:radial}.

In the next step, for each considered radius, these parameters are input for the radiative transfer code \texttt{TITAN}, which calculates the emergent spectrum from an ionized, illuminated, and dissipated atmosphere, i.e., optically thick warm corona 
with a defined optical thickness $\tau_{\rm WC}$. For this paper, we are interested in the iron line spectral region; therefore, we treat Compton scattering only in the expression of cooling and heating functions as described in Sec.~\ref{sec:titan} below. The full code, which solves photoionization with Compton scattering on the microscopic level, is under construction by our team, and its results will be presented in our future work. 

Finally, after collecting the angle and radial dependent intensity field for a given global disk parameters, as: 
$M$, $a$, and $\lambda$, we use \texttt{GYOTO} to calculate the image and spectrum seen by a distant observer (see Sec.~\ref{sec:gyoto}). We analyze the emergent spectra as a function of black hole spin, viewing angle, and disk matter structure for the model setup given in Sec.~\ref{sec:setup}.

\subsection{Radial Stratification}
\label{sec:radial}

\begin{figure*}[!h]
    \centering
    \includegraphics[width=0.99\linewidth]{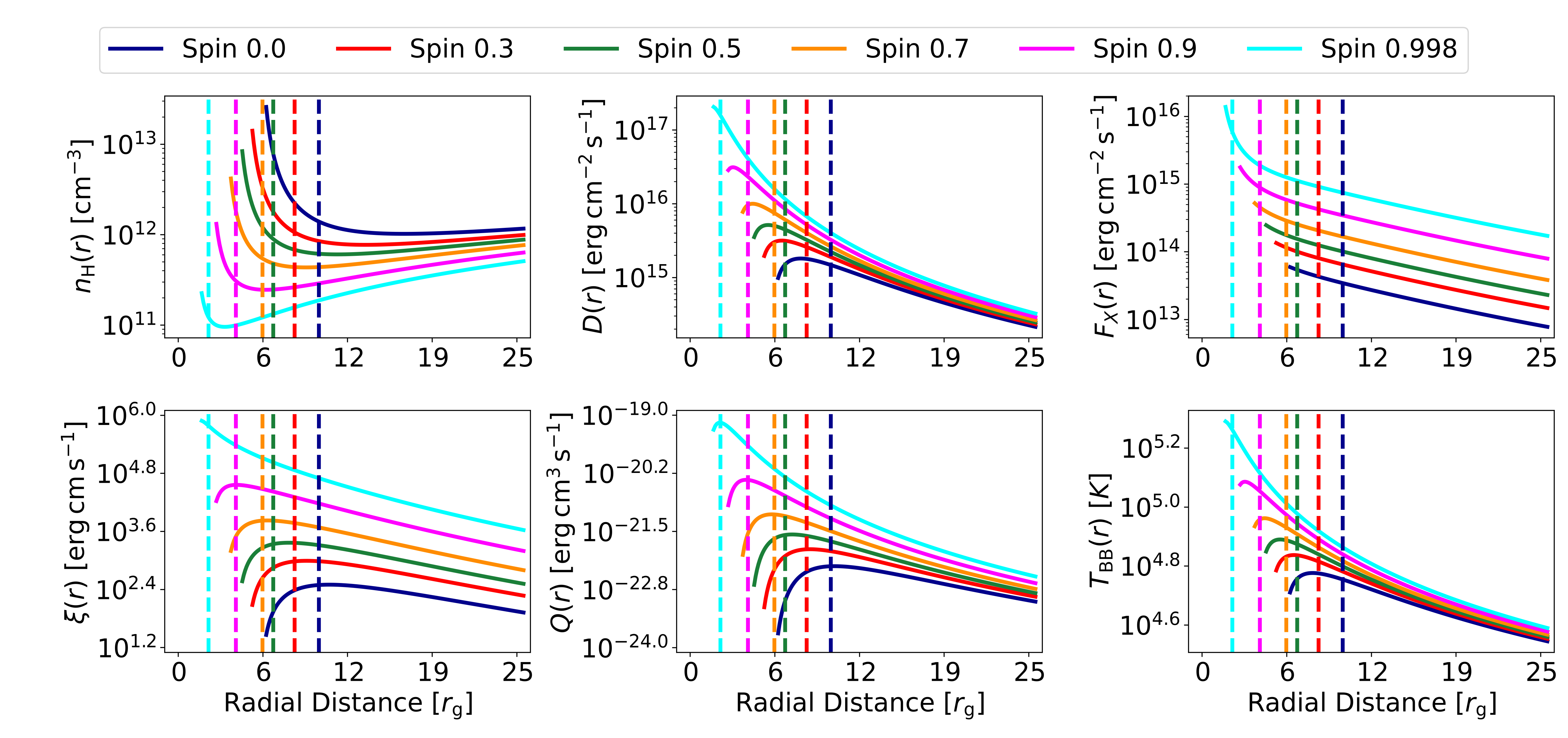}
    \caption{Radial distribution of global model parameters that are inputs for \texttt{TITAN} code. The colors represent different spin values given in the top box. The subplots are as follows: top left panel: gas number density [Eq. \ref{eq:nh_r}], top middle panel: dissipation flux [Eq. \ref{eq:D_r}], top right panel: X-ray flux from a hot corona [Eq. \ref{eq:FX_r}], bottom left panel: ionization parameter [Eq. \ref{eq:xi_r}], bottom middle panel:  internal heating of the warm corona [Eq. \ref{eq:Q_r}], and  bottom right: accretion disk temperature $T_{\rm BB}$ [Eq. \ref{eq:T_bb_r}].
   The vertical dashed lines mark the radius at which the output temperature structure from \texttt{TITAN} reaches the maximum, see Fig.~\ref{fig:matter_structure} and text for details. The computational parameters are described in Sec.~\ref{sec:setup} and Tab.~\ref{tab:param_space}, with $M$, $h$, $f_X$, and ${ f_{W}}$ kept constant.}
    \label{fig:mod_params}
\end{figure*}

For the radial stratification of the warm corona, we follow the same prescription as in \citet{2022MNRAS.515..353X}; hence, in the following section, we reiterate the formalisms they described.

In our approach, the disk is radially stratified into 20 uniformly spaced zones from $r_{\rm in} = r_{\rm ISCO} + 1\, r_{\rm g}$ to $r_{\rm out}= 25\, r_{\rm g}$ (we assume that the maximum flux comes from within this region and the effects of strong gravity is minor beyond this radii)
(see Appendix \ref{Appx:cold_line}). Here, ISCO is defined as the innermost stable circular orbit of a prograde accretion disk, and $r_{\rm g} = GM/c^2$ is the gravitational radius defined by the mass $M$ of the black hole, with $G$ and $c$ being the gravitational constant and the speed of light in vacuum, respectively. Note that $1\, r_{\rm g}$ is added to avoid the nonphysical rise of the disk number density in Eq.~\ref{eq:nh_r} \citep{2022MNRAS.515..353X}. 

To compute the radially stratified gas number density of the warm corona, we first consider the mid-plane density from \citet{1994ApJ...436..599S} and assume the density falls by $10^3$ to the surface \citep[e.g.][]{2019MNRAS.489.3436J}. Hence, the gas number density in 
cm$^{-3}$, for a disk atmosphere, is given by
\begin{equation}
    n_{\rm H} = \left( 2.4 \times 10^{15}\right) \left( \frac{\eta}{0.1}\right)^2 \left( \frac{\alpha}{0.1}\right)^{-1} \left( \frac{M}{\rm M_{\odot}}\right)^{-1} \times \lambda^{-2} r^{3/2} J(r)^{-2}\ ,
    \label{eq:nh_r}
\end{equation}
where $r$ is the radial distance in units of $r_{\rm g}$, the accretion rate $\lambda = L_{\rm bol}/L_{\rm Edd}$ 
 where $L_{\rm bol}$ is the bolometric luminosity and $L_{\mathrm{Edd}} = {\mathbf 4 \pi G M m_{\rm p} c/\sigma_{\rm T}}$ is the Eddington luminosity {(with $\mathbf{m_{\rm p}}$ being mass of proton and ${\rm \sigma_{\rm T}}$ as the Thompson cross section)}, $\eta$ is the accretion efficiency, $\alpha$ is the viscosity parameter, and $J(r) = 1-(r_{\rm ISCO}/r)^{1/2}$ 
 is the stress-free inner boundary condition in a standard thin disk model \citep{1973A&A....24..337S}. The radial dependence of the gas number density is depicted in the top left panel of Fig.~\ref{fig:mod_params}.

The total energy flux $D(r)$ in erg\,cm$^{-2}$\,s$^{-1}$, produced by each side of the standard accretion disk due to dissipation, is described by
\begin{equation}
    D(r) = \left( 6.89 \times 10^{27} \right) \left( \frac{\eta}{0.1} \right)^{-1} \left( \frac{M}{\rm M_\odot} \right)^{-1} \lambda r^{-3} J(r)\ .
    \label{eq:D_r}
\end{equation}
This dissipative flux is further divided into three contributing fractions: illuminating X-ray flux fraction $f_X$ determines the total X-ray luminosity of the hot corona, i.e. lamp; the dissipative flux fraction ${ f_{W}}$ determines the dissipation due to internal heating of the disk atmosphere, i.e. warm corona; and $(1 - f_X - { f_{W}})$ of $D(r)$ determines the accretion disk flux that enters the atmosphere from below in the form of black body radiation with $T_{\rm BB}$ temperature.

For two-corona geometry, we assume a power law spectrum emitted by the hot corona (lamp) with the photon index $\Gamma$ following the relation $E^{-\Gamma} \propto {\rm photon\ flux}$. This hot corona within a critical radius $r_{\rm c}$ emits the X-ray luminosity
in erg\,s$^{-2}$
\begin{equation}
    L_X = \left( 1.5 \times 10^{38} \right) \left( \frac{\eta}{0.1} \right)^{-1} \left( \frac{M}{\rm M_{\odot}} \right) \lambda \int^{r_{\rm c}}_{r_{\rm ISCO}} f_{\rm X} r^{-3} J(r) {\rm d}S_{\rm disk}(r)\ ,
    \label{eq:LX}
\end{equation}
 where ${\rm d}S_{\rm disk}$ is the area of the proper disk element adapted from \citet{2016A&A...590A.132V} that depends on the black hole's spin and radius as
 \begin{equation}
     {\rm d}S_{\rm disk} = 2 \pi r \sqrt{\frac{r^2 + a^2 + 2a^2/r}{r^2 -2r + a^2}} {\rm d}r\ . 
     \label{eq:dS_disk}
 \end{equation}
 The critical radius of the hot corona $r_{\rm c}$ is assumed to be at $10\, r_{\rm g}$, which considers a highly compact corona consistent with observations \citep[e.g.][]{2004MNRAS.349.1435M, 2013ApJ...769L...7R}. 
 Following \citet{2022MNRAS.515..353X}, we assume that the
 regions above $r_{\rm c}$ do not contribute to the hot corona, due to the fact 
 that the dissipation factor falls rapidly beyond $10\, r_{\rm g}$ having negligible effects on the hard X-ray spectrum.

Given that the X-ray luminosity originates from the lamp-post source, the incident X-ray flux reaching the disk atmosphere is subjected to gravitational light bending, and hence depends on the height of the lamp $h$ and the dimensionless spin parameter  $a$ \citep[e.g][]{2004MNRAS.349.1435M,2007ApJ...664...14F,2013MNRAS.430.1694D}. The radially stratified X-ray flux in erg\,cm$^{-2}$\,s$^{-1}$, is defined as \citep{2017MNRAS.472L..60B}
\begin{equation}
    F_{X}(r) = \frac{L_X F(r,h) g_{\rm lp}^2}{z(M) A}\ ,
    \label{eq:FX_r}
\end{equation}
where $F(r,h)$ is given by the fitting formula proposed by \citet{2007ApJ...664...14F} for the illumination pattern of the disk, $g_{\rm lp}$ is the ratio of the photon frequency on the disk to that at the X-ray source \citep{2013MNRAS.430.1694D},
\begin{equation}
    g_{\rm lp} = \frac{r^{3/2} + a}{\sqrt{r^2 + 2ar^{3/2} - 3r^2}}\sqrt{\frac{h^2 + a^2 -2h}{h^2 + a^2}}\ ,
    \label{eq:glp}
\end{equation}
$z(M)$ is used to convert the area into physical units \citep{2017MNRAS.472L..60B}
\begin{equation}
    z(M) = \left( \frac{G{\rm M_{\odot}}}{c^2} \right)^2 \left( \frac{M}{{\rm M_{\odot}}}\right)^2\ ,
    \label{eq:z_M}
\end{equation}
and the normalization factor
\begin{equation}
    A = \int^{r_{\rm out}}_{r_{\rm ISCO}} F(r,h) g_{\rm lp}^2 {\rm d}S(r)
    \label{eq:Norm_Lumin}
\end{equation}
is to make sure the integration of the total flux equals $L_X$.

The X-ray flux and the gas  number density can further be used to define the ionization parameter $\xi(r)$ in ${\rm erg\, cm\, s^{-1}}$ as
\begin{equation}
    \xi(r) = \frac{4 \pi F_{X}(r)}{n_{\rm H}(r)}\ .
    \label{eq:xi_r}
\end{equation}
The radial dependence of the X-ray flux and the ionization parameter is depicted in Fig.~\ref{fig:mod_params}.

The internal heating $Q(r)$ in units of ${\rm erg\, cm^2\, s^{-1}}$ occurring in the disk atmosphere is uniformly distributed across the vertical structure of the warm corona with the given $\tau_{\rm WC}$, and it is defined by the dissipative flux fraction  ${ f_{W}}$ as
\begin{equation}
    Q(r) = \frac{{ f_{W}} D(r) \sigma_{\rm T}}{\tau_{\rm WC} n_{\rm H}(r)}\ .
    \label{eq:Q_r}
\end{equation}

Finally, the remaining dissipative energy contributes to the warm corona as a back-illumination from the accretion disk. It enters an atmosphere from below and follows the black body shape of the spectrum with the corresponding temperature $T_{\rm BB}(r)$ in K given by:
\begin{equation}
    T_{\rm BB}(r) =  \left(\left(1 - { f_{W}} -f_X\right)\frac{D(r)}{\sigma} \right)^{1/4}\ ,
    \label{eq:T_bb_r}
\end{equation}
where $\sigma$ is the Stefan-Boltzmann constant.
The above radially stratified parameters are further used as the input parameters to the photoionization code \texttt{TITAN}, described in the section below.

\subsection{Radiative transfer with \texttt{TITAN}}
\label{sec:titan}

The radiation emitted from the atmosphere covering the underlying accretion disk is computed using the photoionization code \texttt{TITAN}, specifically with its most advanced version described by \citet{2003A&A...407...13D}. 
With the given input parameters, \texttt{TITAN} considers a plane-parallel slab of gas and solves the radiative transfer
\begin{equation}
    \mu \frac{{\rm d}I_{\nu}}{{\rm d}\tau_\nu} = -I_{\nu} + S_{\nu},
    \label{eq:rad_trans}
\end{equation}
where $\mu$ is the cosine of the angle between the normal and the light ray, $I_{\nu}$ is the specific intensity at the angle $\mu$, $S_{\nu}$ is the source function, and $\tau_{\nu}$ is the optical depth.
The slab is divided into discrete horizontal layers, and the transfer of radiation is calculated using the Accelerated Lambda Iteration method under non-Local Thermodynamic Equilibrium (nLTE) conditions \citep{2003A&A...407...13D}. It assumes a two-stream approximation where the two streams are the intensities computed from top to bottom of the disk atmosphere as the inward flux and from bottom to top as the backward flux. 

Within each layer, the physical state of the gas (ion abundances, temperature, and level population) is computed by solving a local balance between the ionization and recombination ions, excitations and de-excitations, local energy balance (radiative and mechanical heating is balanced by radiative cooling), and total energy balance (equality of total inward and outward energy flows) \citep{2000A&A...357..823D}. The local energy balance equation includes all relevant radiative processes such as: free-free processes, i.e., bremsstrahlung; Compton heating and cooling; bound-free processes, i.e., photoionization and recombination; and finally bound-bound processes, such as line heating and cooling.

The number of ionic transitions taken into account is 4141 in the energy range between 0.01 eV and 25 keV. The ten most abundant elements are taken into account: H, He, C, N, O, Ne, Mg, Si, S, and Fe with Solar abundances.
In general, it is possible to change gas content, but due to the model complexity, we keep them constant at the values provided by \citet{grevesse89}.
There are two more parameters, which are an input to \texttt{TITAN}'s models, but we keep them constant for better visualization of our results. Those parameters are: the total column density $N_{\rm H}$, and the photon index $\Gamma$ of the power-law shape of incident hard X-rays (see Sec.~\ref{sec:setup} for their values).

Currently, \texttt{TITAN} has two modes of operation, with constant density and with constant total pressure. The constant density mode assumes a constant density for all the layers of the slab/disk atmosphere, which are stratified against temperature. While the constant pressure assumes a constant total (i.e., gas + radiation) pressure across the structure, varying both density and temperature. For this paper, we have prescribed a constant density, where at each radius the density number $n_{\rm H}(r)$ is given by Eq.~\ref{eq:nh_r}. The rest of the input parameters required by the \texttt{TITAN} code are: ionization parameter $\xi(r)$, internal heating of the warm corona $Q(r)$, 
and black body temperature $T_{\rm BB}(r)$, and they change with radius 
according to radially stratified Eqs.~\ref{eq:xi_r},~\ref{eq:Q_r}, and 
\ref{eq:T_bb_r}, respectively.

\begin{table}[h]
    \begin{center}
    \caption{Angular description of directions used in \texttt{TITAN}'s output. The first column shows the angle $i$ between the slab’s normal and the observer’s line of sight, which is fixed in \texttt{TITAN}. The second column gives the cosine of angle $i$, and the third column lists the corresponding normalized solid angle.}
    \begin{tabular}{ccc}
        \hline
       emission angles (i) & $\cos(i)$ & $\delta\Omega/4\pi$ \\
        \hline
        \hline
        7' & 1 & 1e-6 \\ 
        $17.6^{\circ}$ & 0.9531 & 0.0592 \\ 
        $39.7^{\circ}$ & 0.7692 & 0.1197 \\
        $60^{\circ}$ & 0.5 & 0.1422 \\
        $76.7^{\circ}$ & 0.2308 & 0.1197 \\
        $87.3^{\circ}$ & 0.0469 & 0.0592 \\
        \hline
    \end{tabular}
    \label{tab:solid_angle}
    \end{center}
\end{table}

\texttt{TITAN} computes the output intensity spectra for 6 fixed angles $i$, listed in Tab.~\ref{tab:solid_angle}, defined as an inclination between the normal to the slab and the line of sight towards an observer. In addition, Tab.~\ref{tab:solid_angle} displays the cosine of these angles in column 2, and solid angles being roots of the Gauss-Legendre quadrature used for 
numerical integration of total energy flux. These angle-dependent intensities are further used as input parameters for the ray-tracing code \texttt{GYOTO}. 
Note that the above angles are used only to represent the angular dependence of local intensities emitted from the slab, and the proper viewing angle of the total system is an output parameter given by  \texttt{GYOTO} during ray-tracing computations.

\subsection{Relativistic ray-tracing with \texttt{GYOTO}}
\label{sec:gyoto}

\texttt{GYOTO} is a relativistic ray-tracing code that computes null geodesics for photons using the \(3+1\) formalism, in which a four-dimensional spacetime is foliated into three-dimensional spatial hypersurfaces and a one-dimensional timelike direction.
\citep[see][]{2007gr.qc.....3035G,  2008itnr.book.....A, 2010nure.book.....B}. It assumes a screen with a resolution of $N = n\times n$, where $N$ is the total number of pixels and $n$ is the number of pixels on one side of the square screen. Hence, a total of $N$ photon paths are traced back to the source -- one from each pixel on the screen
 \citep{1979luminet}.

To compute the ray-traced spectrum with \texttt{GYOTO}, we used the Directional Disk option of the code, which was previously tested by \citet{2016A&A...590A.132V} in the case of an illuminated accretion disk around a stellar mass black hole. This model allows us to include the relativistic effects caused by a central black hole on the intensity spectrum emitted locally from a flat equatorial disk. The spectral input is organized as intensity varying with both viewing angle and radial distance from the black hole, while the disk is cylindrically symmetric and emits only from its surface. The model is constructed under the assumption of the Kerr geometry, assuming Boyer-Lindquist coordinates; thus, both the central black hole’s mass and spin are also input parameters.

The energy dependent intensities from \texttt{TITAN} are treated as the emitted intensities $I_{\nu_{\rm em}}$, while the corresponding intensities measured by a distant observer in erg\,cm$^{-2}$\,s$^{-1}$\, Hz$^{-1}$\,sr$^{-1}$ are given by:
\begin{equation}
    I_{\nu_{\rm obs}} = g^3 I_{\nu_{\rm em}}\,,
\end{equation}
where $g = \nu_{\rm obs} / \nu_{\rm em}$ is the redshift factor, with $\nu_{\rm obs}$ and $\nu_{\rm em}$ as the observed and emitted photon frequency, respectively.
The computed observed specific intensity is then further used to calculate the observed monochromatic flux in erg\,cm$^{-2}$\,s$^{-1}$\, Hz$^{-1}$:
\begin{equation}
    {\rm d}F_{\nu_{\rm obs}} = I_{\nu_{\rm obs}} \cos(\theta_\mathrm{obs}) {\rm d}\Omega_{\rm obs},
    \label{eq:dflx_gyoto}
\end{equation}
subtended by the solid angle ${\rm d}\Omega_{\rm obs}$, where $\theta_{\rm obs}$ is the angle between the normal to the screen and the normal to the accretion disk, which corresponds to the viewing angle. 
As customary in ray tracing calculations, the observed solid angle equals the 
angle that subtends the field of view used on the screen.

Out of the total number of $N$ pixels on the screen, each pixel (assumed to be a point-like object) corresponds to a specific direction in the sky. The object on the screen is defined by the field of view, the pixels that cover the field of view, and the spectral property used, such as the channels and their corresponding wavelengths (or energies) \citep{2011CQGra..28v5011V}. A small field of view $\delta \Omega_{\rm px}$ is covered by each pixel, which is defined as the total field of view divided by the total number of pixels
\begin{equation}
    \delta \Omega_{\rm px} = \frac{2\pi\, (1-\cos f)}{N}, \
\end{equation}
where $f$ is the angle of the normal to the screen and the most distant incident angle of the photons on the screen. Using Eq.~\ref{eq:dflx_gyoto}, the total observed flux, hence, can be defined by
\begin{equation}
    F_{\nu_{\rm obs}} = \sum_{N} I_{\nu_{\rm obs},{\rm px}} \, \cos(\theta_{\rm px})\, \delta \Omega_{\rm px}\, ,
\end{equation}
where $\theta_{\rm px}$ is the angle between the normal to the screen and the direction of incidence corresponding to this pixel. $\theta_{\rm px}$ is not identical to $\theta_{\rm obs}$, as the screen is not a point source. At large observer(screen)–source distances, the angle subtended by a pixel, $\theta_{\rm px}$, may differ only slightly from the viewing angle $\theta_{\rm obs}$; however, at shorter distances, this deviation can become significant. At each frequency, the flux at every pixel is computed based on the radial distance and viewing angle corresponding to that pixel, resulting in a two-dimensional image. Integrating over this image gives the total flux at that frequency. Repeating this process across all frequencies yields the relativistically corrected spectrum. The detailed computations of the frequency-dependent flux with more models can be found in \citet{2011CQGra..28v5011V}. 
Finally, the observed flux is transformed into frequency-dependent luminosity using
\begin{equation}
    L_{\nu} = 4 \pi D^2 F_{\nu_{\rm obs}} 
\end{equation}
where $L_{\nu}$ is defined in ${\rm erg \, s^{-1}\, Hz^{-1}}$ and $D$ is the distance of the source from the observer.

\subsection{Setup of  parameters}
\label{sec:setup}

In this paper, we put special attention on the spectral shape of the iron line modified by gravitational effects from the SMBH. Therefore, we test only one case of black hole mass $M = 10^8\, {\rm M_{\odot}}$ and
accretion rate $\lambda = 0.1$, concentrating solely on the dependence of black hole spin -- $a$, viewing angle -- $\theta_{\rm obs}$, 
fraction of X-ray illuminating flux from the lamp -- $f_{X}$, fraction of the dissipation in warm corona -- ${ f_{W}}$, and the height of the lamp -- $h$. The rest of the global disk parameters are taken to be fixed at: $\eta = 0.1$, and $\alpha = 0.1$.
{ We are aware that the accretion efficiency depends on the SMBH spin. But since we are not solving any dynamical equations of an accretion flow, and this value is only needed to define the accretion rate and therefore total energy dissipated at the given radius, we keep it constant for all considered models (also prescribed in \citet{2022MNRAS.515..353X}).}

The variation of model parameters that affect the shape of the outgoing iron line profile is depicted in Table~\ref{tab:param_space}. We show our parameter space in the table with the first column depicting the variable parameter, the second one displays values 
of the computed grid, and the last column shows the canonical parameters, i.e., constant values for each parameter, while others are varied. The canonical values allow us to isolate and study the effect of each parameter individually. The spin dependence occurs in the inner boundary condition $J(r)$, due to the different
$r_{\rm ISCO}$ value. Since the accretion disk is modeled as geometrically thin and optically thick, we assume a constant optical depth of the top layer of the warm corona on the value of \( \tau_{\rm WC} = 5 \) across the entire radial structure. We are aware that some hard X-ray photons are getting thermalized at larger optical depths, but since our conditions change with radius, we keep this amount at a moderate value and uniform across all radially stratified zones. As the incident photon spectrum follows a power-law distribution, we fix the photon index to $\Gamma = 2$ at all radial points. There is a possibility to change the elemental
abundances in \texttt{TITAN} code, but in this paper, we keep them fixed to the Solar values.

\begin{table}[h]
    \begin{center}
    \caption{The list of model parameters that affect the shape of iron line profile: black hole spin - $a$, viewing angle - $\theta_{\rm obs}$, fraction of X-ray illuminating flux from the lamp - $f_{X}$, fraction of the dissipation in warm corona - ${ f_{W}}$, and the height of the lamp - $h$. The first column lists the parameters we explore. The second one displays the values considered for each parameter. The third column provides the canonical values, i.e., the fixed values assigned to all other parameters while a particular parameter is varied. {The models were computed assuming the following parameters being constant: $M = 10^{8}\, {\rm M_{\odot}}$, $\lambda = 0.1$.}}
    \begin{tabular}{ccc}
        \hline
        Variable & Computed & Canonical\\
        parameter & Grid & Value\\
        \hline
        \hline
        & & \\
        $a$ & 0, 0.3, 0.5, 0.7, 0.9, 0.998 & 0.9 \\ 
        & & \\
        $\theta_{\rm obs}$ & 5$^{\circ}$, 10$^{\circ}$, 15$^{\circ}$, 20$^{\circ}$, 25$^{\circ}$, 30$^{\circ}$, & 15$^{\circ}$\\
        & 40$^{\circ}$, 50$^{\circ}$, 60$^{\circ}$, 70$^{\circ}$, 80$^{\circ}$, 85$^{\circ}$ &  \\ 
        & & \\
       $f_X$ & 0.05, 0.1, 0.15,  & 0.2 \\
         & 0.2, 0.25, 0.3   &   \\
          & & \\
        ${ f_{W}}$ & 0, 0.2, 0.4, 0.6, 0.8 & 0.4 \\
        & & \\
        $h$ & 10, 12.5, 15, 17.5, 20 (in ${\rm r_g}$)& 20 ${\rm r_g}$\\
        & & \\
        \hline
    \end{tabular}
    \label{tab:param_space}
    \end{center}
\end{table}

Computations by \texttt{TITAN} result in angle-dependent intensities at each considered radial point. These angles and radial-dependent intensities are then used as the input spectra into  \texttt{GYOTO} code that integrates them over the disk surface, considering all relativistic effects in the Kerr geometry assuming Boyer-Lindquist coordinates. In principle, we can study the final outgoing spectrum for a large grid of parameter space, but for this paper, we consider values given in Tab.~\ref{tab:param_space}. The observer's viewing angle directly corresponds to $\theta_{\rm obs}$ from Eq.~\ref{eq:dflx_gyoto}, and the lowest value reflects face-on projection, while the highest value provides an edge-on view of the source.

The field of view is set to $5\, {\rm \mu as}$ to study the inner region of the SMBH, and the observer is assumed to be at a large distance of $D = 16.9$ Mpc, far enough to ensure flat spacetime. We have kept the resolution of $N=801\times801$ pixels for our simulations, which is reasonable for image resolution in the X-ray domain.

\texttt{GYOTO} can also generate photons originating from the photon ring region, which are classified as higher-order photons due to the multiple orbits they complete around the compact object. However, this region was excluded from the analysis due to resolution limitations, and only first-order photons were considered. Although higher-order photons were neglected in our work, it was demonstrated by \citet{2021MNRAS.504.3424F} that their contribution to the observable iron line profiles is minimal.

\section{Results} \label{sec:results}

Our results are strictly focused on the iron line profile originating from the innermost region of the SMBH. The correctness of our ray-tracing procedure is tested in the case of a single Gaussian line on a flat continuum in 
Appendix~\ref{Appx:single_lin}. Although the luminosity level of a single line test is consistent with the AGN disk luminosity, to have a realistic scenario, we used the ray-tracing code \texttt{GYOTO} on the 
 reflected spectra calculated by the \texttt{TITAN} code that physically changes with radial distance from the SMBH. We present those results in sections below. 

\begin{figure*}[!h]
    \centering
    \includegraphics[width=1.08\linewidth]{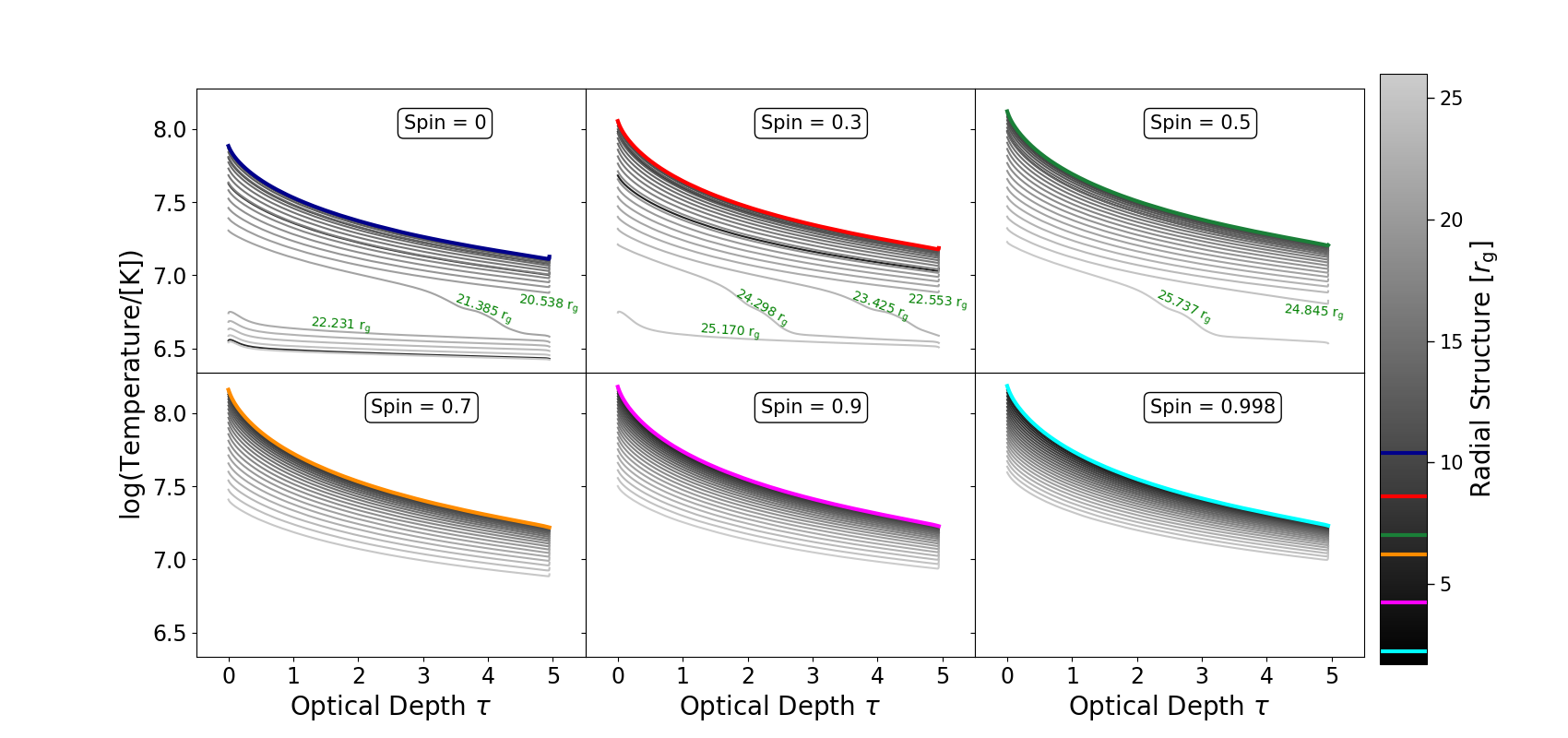}
    \caption{The temperature dependence on the optical depth for various spin values given in the boxes.
    The gray gradient of lines represents the matter structure calculated at different radii, with the darkest line at $r_{\rm in}$, and the lightest line 
    at $r_{\rm out}$. The colored lines show the simulation with maximum temperature, which is coherent with the dashed line in Fig.~\ref{fig:mod_params}. {The corresponding radial point where the temperature drop is observed is { labeled} and marked in green text. The parameters other than spin are chosen to be canonical values from Tab. \ref{tab:param_space}. As this is the temperature structure, it is independent of the viewing angle.}}
    \label{fig:matter_structure}
\end{figure*}

\begin{figure}[!h]
    \centering
    \includegraphics[width=1.0\linewidth]{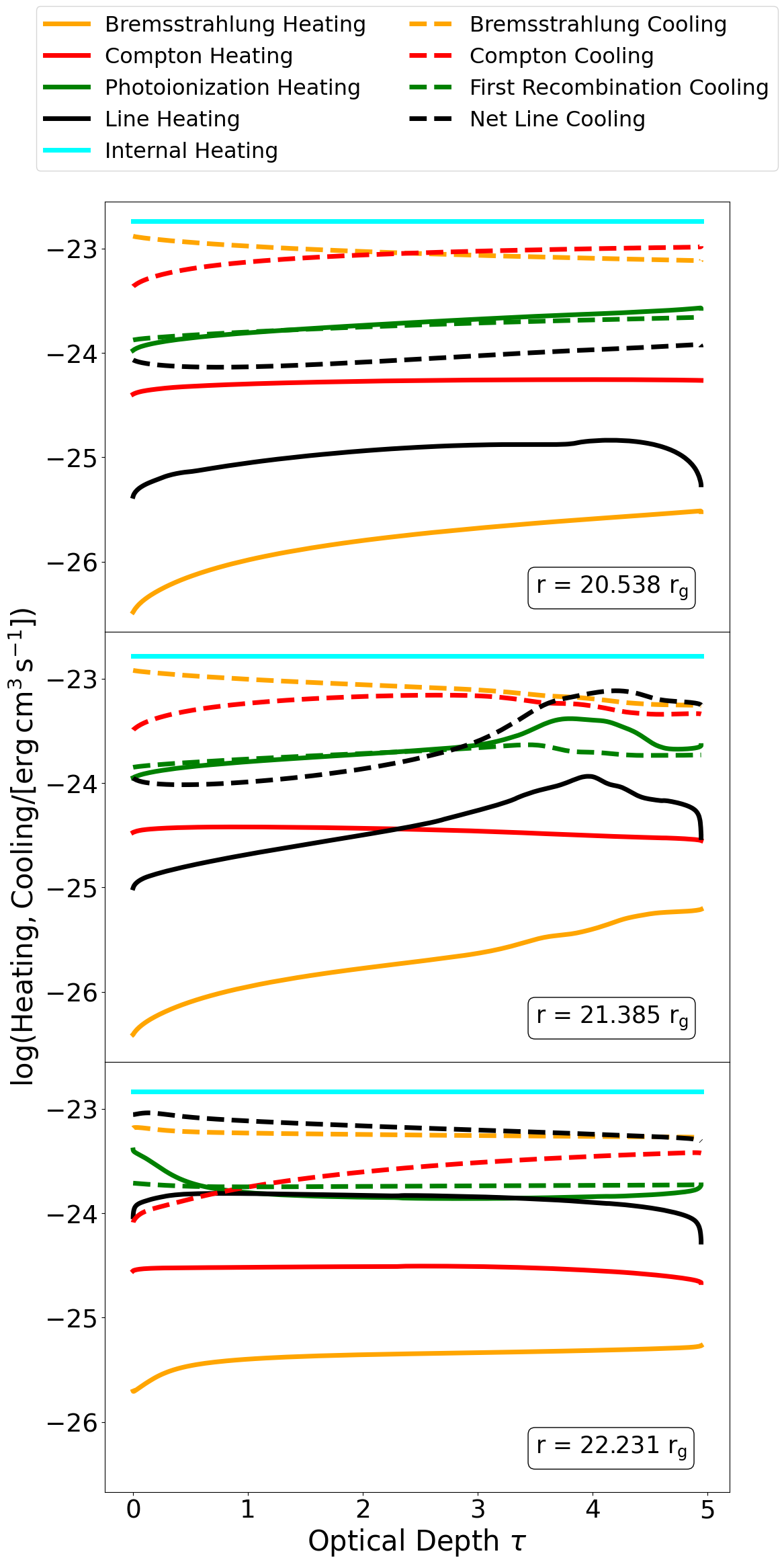}
    \caption{The heating and cooling of the matter structure at equilibrium for spin 0 and mass $10^8 {\rm M_{\odot}}$. The plots depict different points of the radial structure, left: 20.538 $r_{\rm g}$, middle: 21.385 $r_{\rm g}$, right: 22.231 $r_{\rm g}$. The solid lines depict heating and the dashed lines depict cooling. }
    \label{fig:hc_spin_0}
\end{figure}

\begin{figure*}[!h]
    \centering
    \includegraphics[width=1.03\linewidth]{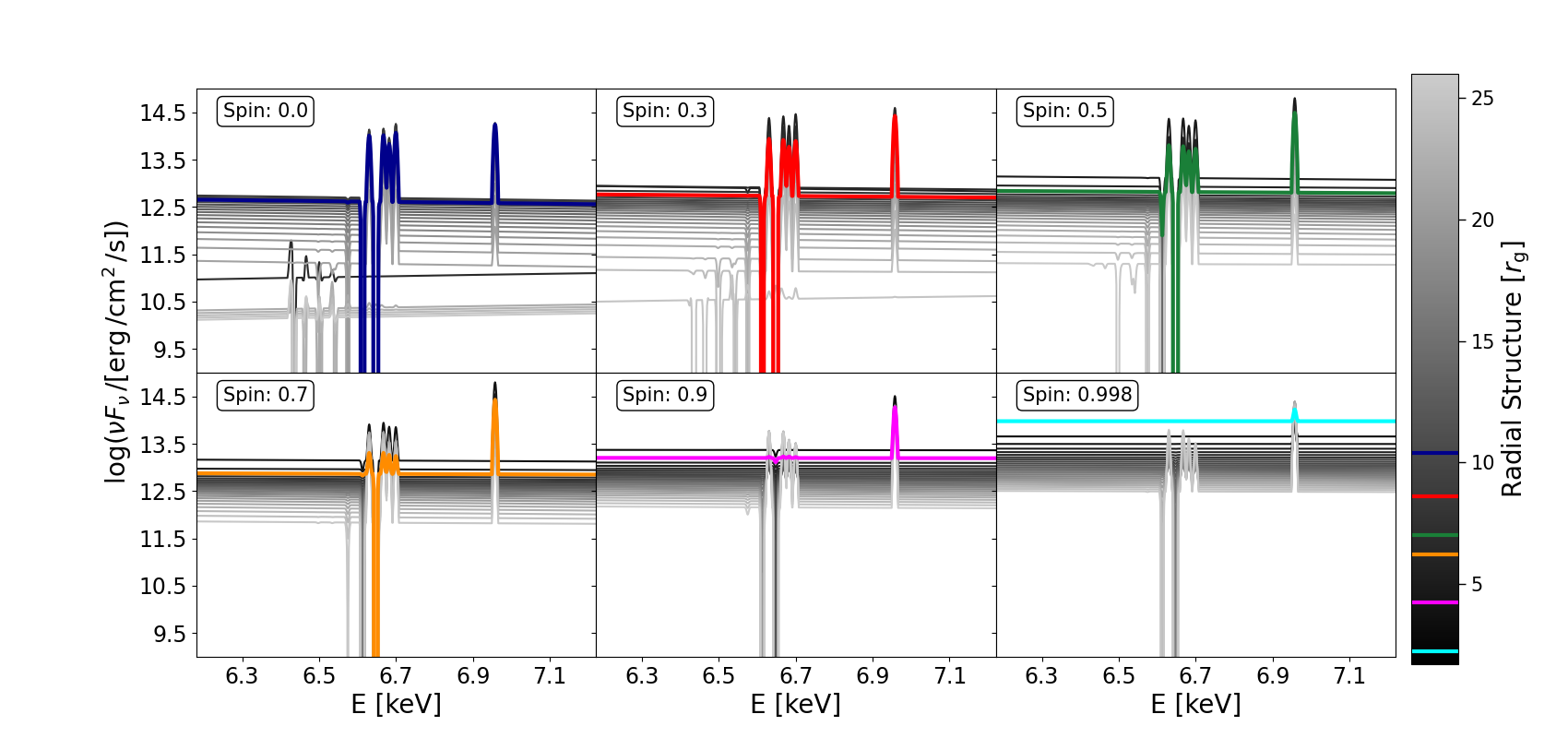}
    \caption{The local spectra $\nu F_{\nu}$ versus photon energy $E$, from \texttt{TITAN} code, for different spin values at the $i = 17.6^\circ$. The gray gradient of lines
    depicts outgoing spectra calculated at different radii,  with the darkest line at $r_{\rm in}$, and the lightest line 
    at $r_{\rm out}$. The colored lines show the simulation with maximum temperature, which is coherent with the dashed line in Fig.~\ref{fig:mod_params} and solid colored lines in  Fig.~\ref{fig:matter_structure}.{The parameters other than the spin and the viewing angle are set to be canonical values as depicted in Tab.~\ref{tab:param_space}. This is the emission spectra, hence the parameter of the angle is chosen from Tab.~\ref{tab:solid_angle}.}}
    \label{fig:spec_spn_0_9}
\end{figure*}

\subsection{Matter Structure and Local Spectra} \label{res:ms}

We use the photoionization code \texttt{TITAN} to generate the local emitted spectrum from the disk atmosphere. At each radially stratified point, \texttt{TITAN} computes the vertical structure, in which the Compton process is simplified, but ionization is computed more rigorously with a large database of atomic transitions. 
The radial evolution of input parameters needed locally for \texttt{TITAN} code,
as described in Section~\ref{sec:radial}, is shown in Fig.~\ref{fig:mod_params}. 
The gas number density is observed to initially fall and then rise with the increase in the radial distance from the SMBH, but in general, the disk atmosphere is very dense, as expected in the case of accretion disks around SMBH. The initial fall occurs due to the stress-free inner boundary condition. 
Hence, very close to $r_{\rm ISCO}$, the gas number density can reach nonphysical values, due to which we start our simulations from $r_{\rm ISCO} + 1\, r_{\rm g}$.

The total energy flux $D(r)$ drops down with $r^{-3}$ dependence, and is presented in the top middle panel of Fig.~\ref{fig:mod_params}. It directly affects the other parameters, such as the X-ray power-law illumination flux $F_{X}(r)$ -- top right panel, the internal heating parameter $Q(r)$ 
-- bottom middle panel and back illumination Temperature $T_{\rm BB} (r)$-- bottom right panel.
Both $Q(r)$ and ionization parameter $\xi(r)$ also exhibit inverse dependencies on the gas number density $n_{\rm H}(r)$ -- specifically, $ Q(r) \propto n_{\rm H}(r)^{-1}$ and $ \xi(r) \propto n_{\rm H}(r)^{-1}$ -- which further suppresses their values. 
With these given input parameters of $n_{\rm H}(r)$, $Q(r)$, $\xi (r)$, and $T_{\rm BB}(r)$ at each 
of the equally spaced 20 radial zones between $r_{\rm ISCO}+ 1 {\rm r_g}$ and 25$r_{\rm g}$, we compute the vertical structure with \texttt{TITAN} radiative transfer code assuming at each radius constant photon index $\Gamma$ and warm corona optical depth $\tau_{\rm WC}$. During radiative transfer computations, we use the assumption of constant density across the vertical gas zones given in the electron scattering optical depth $\tau$.
 
 One of the outputs generated by \texttt{TITAN} is the vertical matter structure with which we can study the temperature at each optical depth of an atmosphere. Such temperature results from the solution of the energy equilibrium equation, where all cooling processes balance all heating processes, radiative plus mechanical.
 The dependence of the temperature vertical profiles for various spin values and at different radial points is depicted in Fig.~\ref{fig:matter_structure}. As we go vertically deeper into the layers, the optical depth increases and the temperature decreases. The temperature also decreases with the increase of the radius, depicted by the change in the gradient of lines, where the dark lines represent the innermost radii and the light gray lines represent the outer regions (see sidebar of the Fig.~\ref{fig:matter_structure}).

As the black hole spin increases, the disk's temperature also increases. This occurs mainly because of the disk model we use, where the amount of energy released increases with spin. The spin affects the inner edge of the disk, and with higher spin, the value of $J(r)$ gets larger. This causes $D(r)$, the energy flux being given off, to increase as well. As a result, the disk gives off more energy overall, making it hotter.

We further detect a sudden fall in the temperature for spin values of: 0, 0.3, and 0.5 at radii 21.385~$r_{\rm g}$, 24.298~$r_{\rm g}$, and 25.737~$ r_{\rm g}$, respectively. This temperature drop may be caused by the change in the radiative heating/cooling mechanism operating in the gas.
To investigate the origin of the above temperature drop, in Fig.~\ref{fig:hc_spin_0} we plot the heating and cooling vertical profiles computed by the \texttt{TITAN} code, for three consecutive
radii: before the drop - upper panel, during the drop - middle panel, and after the drop - bottom panel, for the spin case $a=0$. 
All relevant processes taken into account are listed in the box above the panels. 
We see that in all cases, mechanical internal heating dominates across the atmosphere. Nevertheless, there is a difference between the structure of the radiative cooling, which starts to dominate when the optical depth increases. 
Deeper, at about $\tau=3.5$, net line cooling starts to outweigh Compton and bremsstrahlung cooling, causing the final temperature drop. In the lower panel, when the overall atmospheric temperature is lower, line cooling dominates all cooling processes across the whole thickness of the warm corona. 
 The radial location of the temperature drop corresponding to higher spin shifts outward, implying that line cooling becomes dominant for higher spin at larger radii, i.e., larger than $r_{\rm out}$, outside our computational limit.

The initial parameterization of the input also produces a peak in the temperature profile within the grid domain of our radial structure. In Fig.~\ref{fig:mod_params}, we mark with the dashed lines the radial point where we see the highest temperature in our radial structure. 
Such a peak primarily arises due to the global model, and it closely aligns with the peak of the internal heating parameter $D(r)$.
 This shows that even though roughly 40\% of the dissipative flux is in the form of internal heating, the energy balance at equilibrium is dominated by the heating provided to the structure. Overall, the peak of the matter structure shifts inwards with an increase in spin, suggesting that a higher spin value will have a hotter temperature structure near the ISCO. Since the back-illumination temperature is lower than that of the disk atmosphere, seed photons from the disk in the form of blackbody radiation contribute to the net cooling of the atmospheric layers.

The reflected spectrum produced by \texttt{TITAN} for different spins at an angle of 17.6$^\circ$ is illustrated in Fig.~\ref{fig:spec_spn_0_9}. The direct influence of the temperature can be seen with the variation of the strength of the highly ionized Fe K$\alpha$ lines. With the increase in spin, the temperature of the disk rises, which is apparent from the effect where the FeXXV ions get ionized to FeXXVI, increasing the strength of the line at 6.957 keV. Absorption features are also observed in the spectrum emitted locally from the disk. These arise due to the dissipation of energy within the disk and the presence of photons transmitted from its deeper layers. 

Note that in the case of a warm corona, illuminated photons from the hot corona are generally more energetic than warm gas on the top of the disk and mostly cause photoionization and excitation processes. This gives rise to the line emission, which may exceed the Compton cooling of the matter. On the other side,  soft photons from the disk are the major drivers for Compton up-scattering, i.e,. Compton cooling, which we expect to be responsible for soft X-ray excess emission from the warm corona
\citep{2018A&A...611A..59P,2024MNRAS.530.1603B,2024A&A...690A.308P}. 
Furthermore, in the gas of relatively high density, bremsstrahlung cooling becomes 
dominant, which can provide stability to thermally unstable matter as studied by 
\citet{2019pmdd.book.....A,2023gronkiewicz}. Despite these specific features, we see that under our assumptions, the warm atmosphere can be sustained, and locally the resonant iron K${\alpha}$ lines can be created and then be subject to relativistic blurring. 

\begin{figure}[t]
    \centering
`    \includegraphics[width=0.95\linewidth]{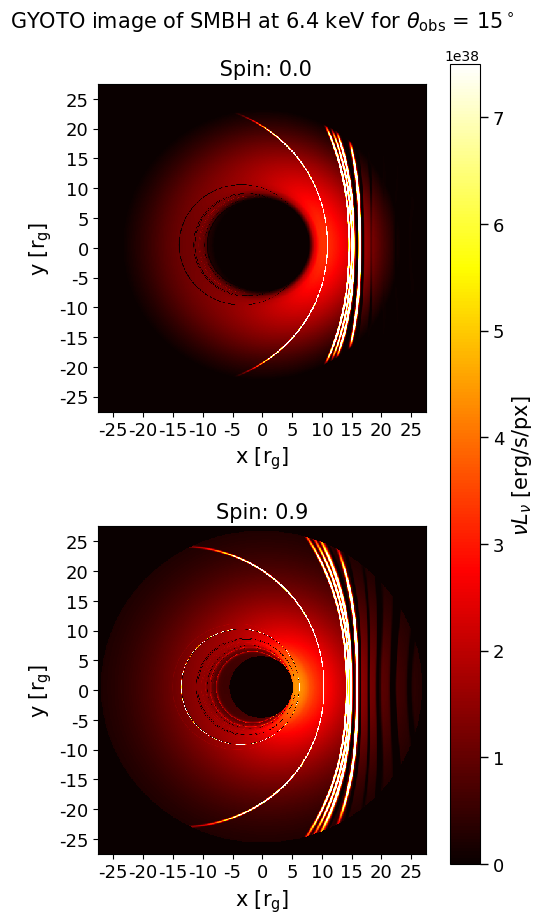}
    \caption{Black hole disk images computed by \texttt{GYOTO} from the \texttt{TITAN} local angle dependent spectra, for two different spins. Images are computed at 6.4~keV and at $\theta_{\rm obs} = 15^\circ$ -- almost face-on. The black hole mass is set to $10^{8}\, {\rm M_{\odot}}$. The map is created for monochromatic luminosity $\nu L_\nu$ in erg~s$^{-1}$~px$^{-1}$, 
     and the X and Y axes are in units of ${\rm r_g}$ with the SMBH center as the origin.}
    \label{fig:img_spn_0_9}
\end{figure}

\begin{figure}[t]
    \centering
    \includegraphics[width=0.95\linewidth]{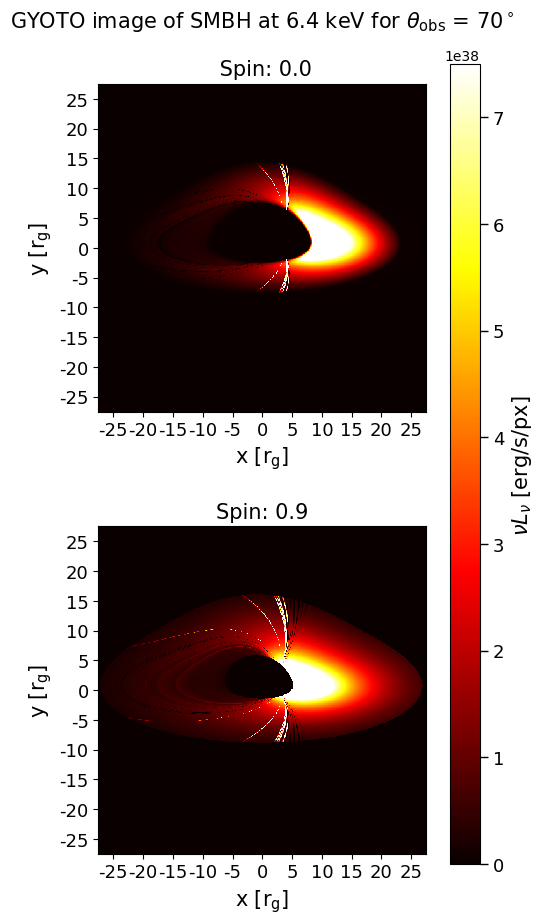}
    \caption{Same as in Fig.~\ref{fig:img_spn_0_9} but for $\theta_{\rm obs} = 70^\circ$ - almost edge on.}
    \label{fig:img_ang_85}
\end{figure}

\subsection{Ray-traced Spectrum} \label{res:spectrum}

In the previous section, we demonstrated that, for the assumptions of the global disk model and with the assumed emitting surface, highly ionized Fe K$\alpha$ lines can be created in the local spectrum of the warm and dissipative corona. This section discusses the outputs from \texttt{GYOTO}, where we include the relativistic effects in the emitted spectrum. We start with an illustration of the image produced by \texttt{GYOTO} at 6.4 keV in Sec.~\ref{sec:gyoto_image}. Following a 2-D integration of the image at each frequency, we obtain the relativistically corrected spectrum. Hence, in the following subsections, we show the dependence of the output spectrum on the spin of SMBH, viewing angle, height of the lamp, and the dissipative flux distribution.

\subsubsection{\texttt{GYOTO} Image} \label{sec:gyoto_image}
 
For the given viewing angle and spatial resolution, the \texttt{GYOTO} code 
produces ray-traced images at different energies. 
The examples of such images for two extreme spin parameters 0 and 0.9, as seen by a distant observer, are presented in Fig.~\ref{fig:img_spn_0_9} in the case 
of the source seen face-on. The influence of the gravitational mass of SMBH is seen as the relativistic distortion of the luminosity emission.

\begin{figure*}[!h]
    \centering
    \includegraphics[width=1.08\linewidth]{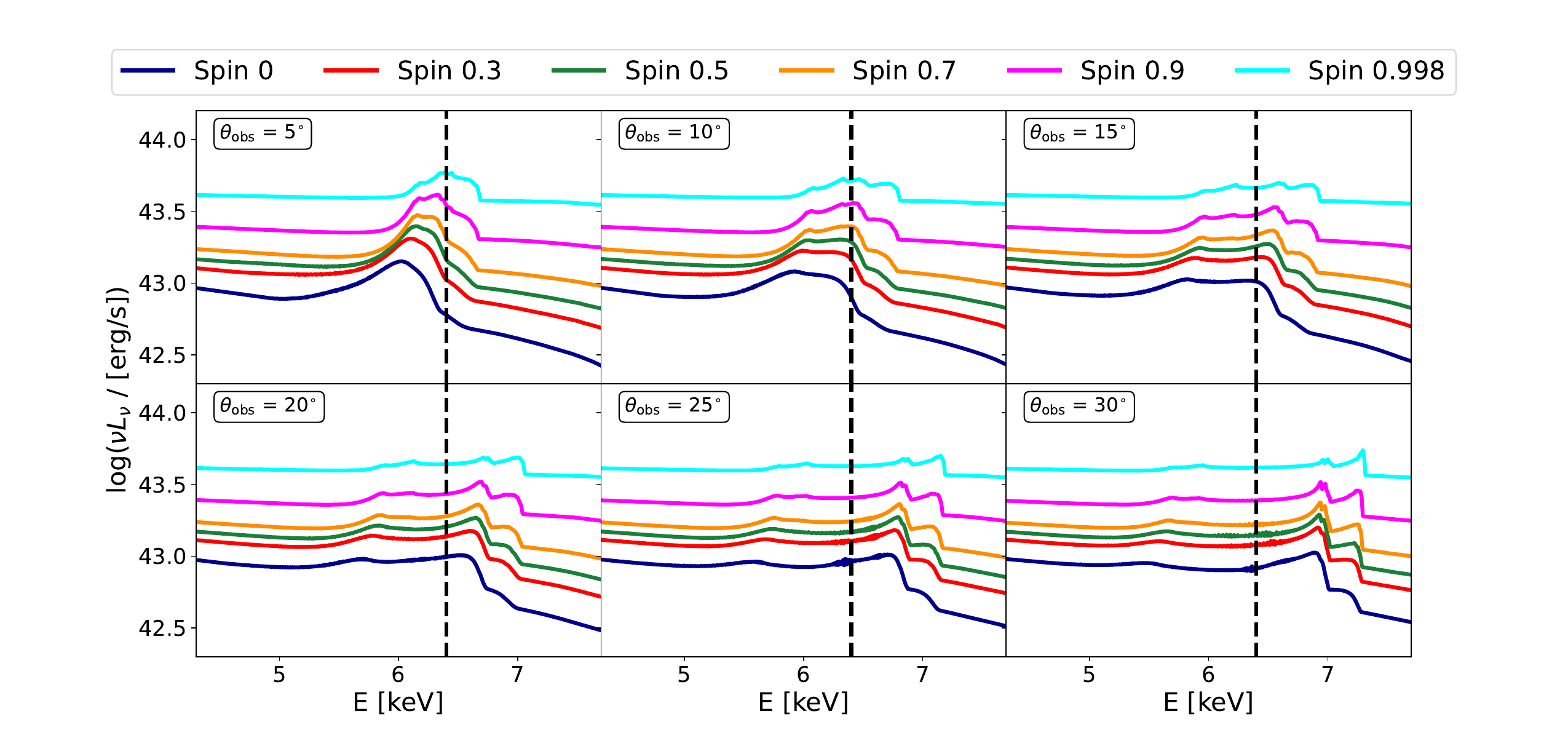}
    \caption{The reflected ray-traced spectra from \texttt{TITAN/GYOTO} computations, for viewing angles ($\theta_{\rm obs}$) of the range from $5^{\circ}$ to $30^{\circ}$ given in the panel's boxes. For each $\theta_{\rm obs}$, different line colors are plotted according to the spin values, displayed in the box above the figure. All spectra are presented by luminosity $\nu L_{\nu}$ versus photon energy $E$.  The vertical dashed line marks the position of 6.4 keV energy.}
    \label{fig:spec-all}
\end{figure*}

\begin{figure*}[!h]
    \centering
    \includegraphics[width=1.08\linewidth]{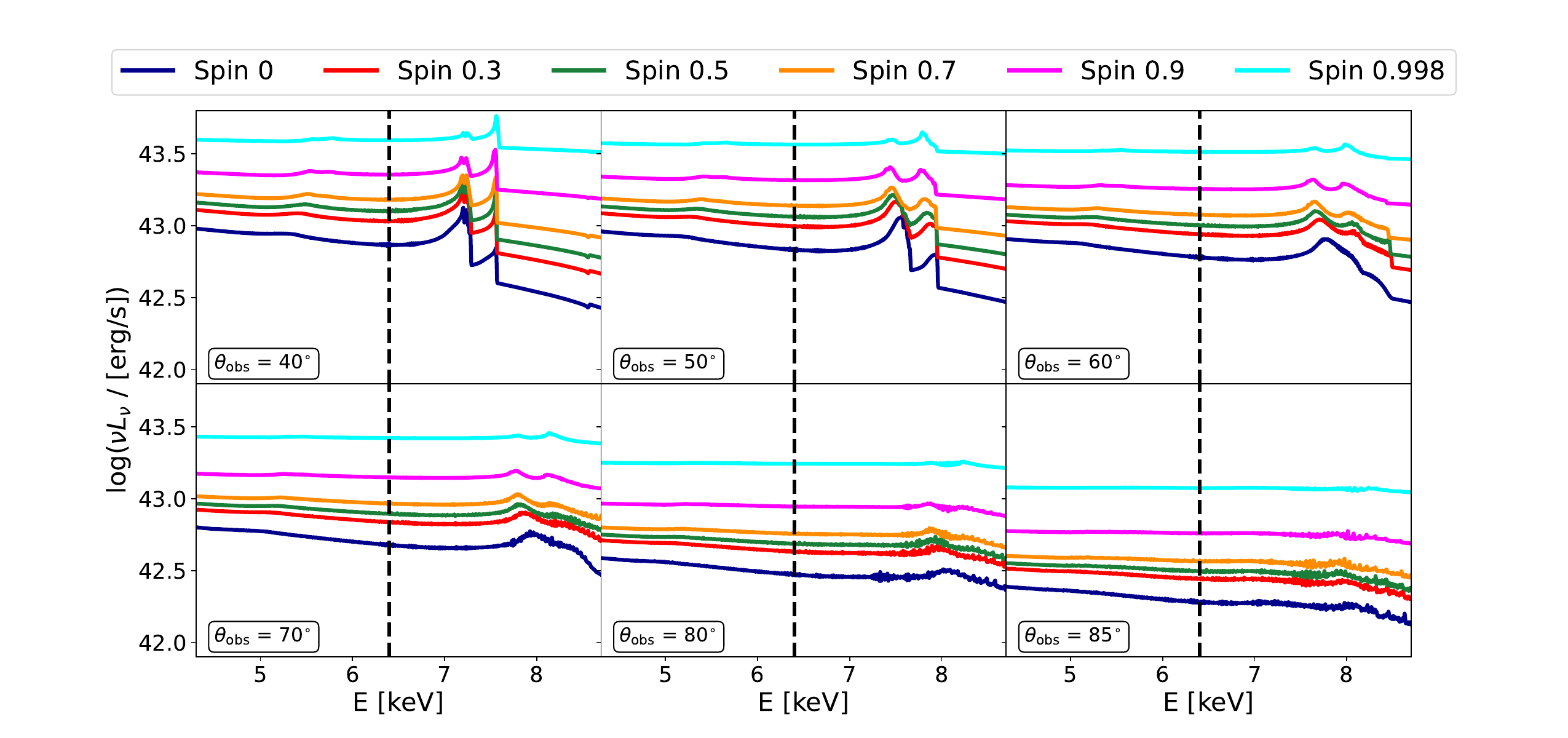}
    \caption{Same as in Fig.~\ref{fig:spec-all}, but for $\theta_{\rm obs}$ for the range from $40^{\circ}$ to $85^{\circ}$ given in the panel's boxes. }
    \label{fig:spec_spn_ang_rest}
\end{figure*}

On the quite uniform luminosity emission map, we also see a few bright and dark stripes in the image. As \texttt{GYOTO} includes all the relativistic effects, the resonant emission line at 6.957 keV from FeXXVI would have to be emitted at a specific distance $r$ from the SMBH center, to be redshifted to the 6.4 keV region. The bright curve close to the SMBH in Fig.~\ref{fig:img_spn_0_9} is the 6.957 keV emission, which shifts to the 6.4 keV region. Similarly, the dark and bright lines show the spectrum's other absorption and emission features moved to the 6.4 keV region. If we compare the spectrum plotted in Fig.~\ref{fig:spec_spn_0_9} to the image produced by \texttt{GYOTO} in Fig.~\ref{fig:img_spn_0_9}, we see a one-to-one correlation of the location of the emission and absorption features, illustrating the gravitational redshift map computed by \texttt{GYOTO}. The curves follow the contours of a constant redshift factor $g$ for the equatorial thin disk.

We also present an almost edge-on accretion disk in Fig.~\ref{fig:img_ang_85}. 
The bright elongated region observed on the right side of an accretion disk is primarily due to the relativistic Doppler boosting, where emission from the approaching side of the disk is enhanced (the disk rotates clockwise as viewed by a face-on observer). In addition, general relativistic light bending allows photons emitted from the other side of the disk, behind the black hole, to be redirected toward the observer. However, the flux observed from this lensed region is significantly lower than the direct disk emission, as only a small fraction of photons are bent along paths that reach the observer at a given $\theta_{\rm obs}$.

\subsubsection{The influence of Viewing Angle and Spin} \label{sec:Spec_spin_ang}

To further study the various scenarios, we present spectra of iron line regions for different spins ($a$) and viewing angles ($\theta_{\rm obs}$) in Figs.~\ref{fig:spec-all} and~\ref{fig:spec_spn_ang_rest}, respectively. In all those simulations, the iron line complex is self-consistently given by photoionization modeling with \texttt{TITAN} radiative transfer code.
  The effective emitting area of the disk increases with increasing black hole spin, since $r_{\rm ISCO}$ moves inwards for higher spin values, as illustrated in  Fig.~\ref{fig:img_spn_0_9}. This trend directly affects the total luminosity, which is reflected in the position of the spectral continuum: for zero spin, the continuum strength is weakest, and it progressively increases with spin as seen in both Figs.~\ref{fig:spec-all} and~\ref{fig:spec_spn_ang_rest}.
 
 The dashed lines in Fig.~\ref{fig:spec-all} and Fig.~\ref{fig:spec_spn_ang_rest} highlight the 6.4 keV energy in the observer's frame. 
  We do not see a neutral emission line for the initial data generated from \texttt{TITAN} in most of our scenarios, with a few exceptions at the outer edge of our radial structure. The major contribution to the observed 6.4 keV spectral feature comes from the Fe K$\alpha$ emission lines of FeXXV and FeXXVI ions generated at 6.63, 6.67, 6.682, 6.7 keV for FeXXV and 6.957 keV for FeXXVI (see Sec.~\ref{subs:litr} for details). 
The high temperature of the matter structure is adequate for the existence of the highly ionized iron ions, while neutral iron does not exist under those physical conditions. Due to the close vicinity to the SMBH, the emission from the highly ionized iron ions gets gravitationally redshifted. For angles below $20^{\circ}$, we observe that the lines are getting redshifted around the 6.4 keV region. In contrast, for $\theta_{\rm obs}$ greater than $20^{\circ}$, the line shift prominently moves out of this region, as seen on the bottom panels of Fig.~\ref{fig:spec-all} and all panels of Fig.~\ref{fig:spec_spn_ang_rest}. 

\begin{figure}[!h]
    \centering
    \includegraphics[width=1.05\linewidth]{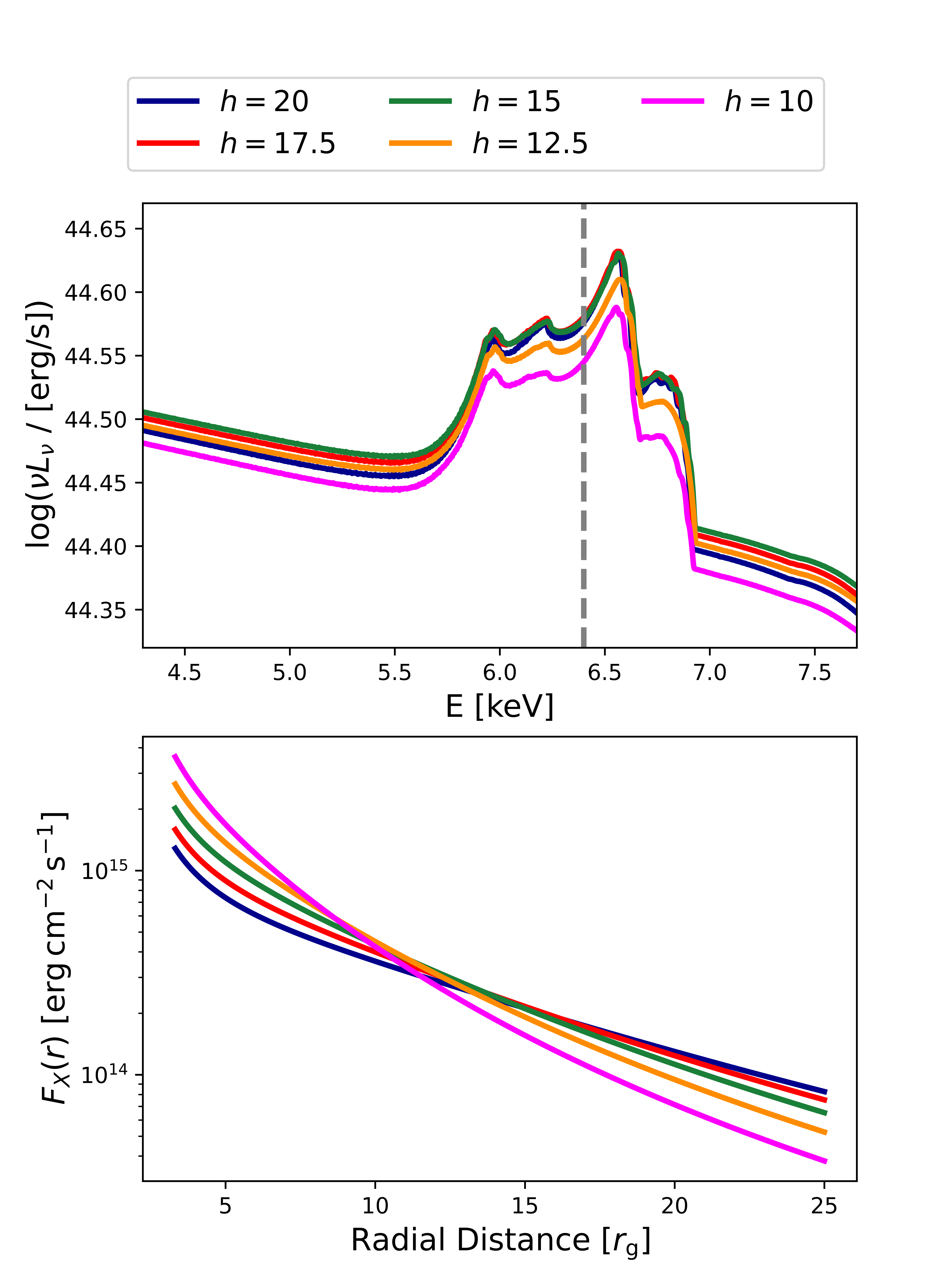}
    \caption{The top panel shows the variation of the Fe K$\alpha$ profile with the lamp height. The other parameters are set to their canonical values as shown in Tab.~\ref{tab:param_space}. The bottom panel shows the illuminating X-ray flux radial profiles,  varying with the lamp height. The height is in units of ${\rm r_g}$ given in the box above panels.}
    \label{fig:height_var}
\end{figure}

The motion of matter in an accretion disk around a SMBH introduces Doppler broadening and blue boosting of the emission lines. For $\theta_{\rm obs}$ beyond $20^{\circ}$, the lines become extremely broadened due to the disk motion, and at a maximal angle of $85^{\circ}$, the spectral features disappear in the underlying continuum, as presented at the bottom right panels of Fig.~\ref{fig:spec_spn_ang_rest}. We also observe a noisier spectrum in Fig.~\ref{fig:spec_spn_ang_rest}. Since \texttt{GYOTO} traces rays backward from the screen to the disk, some rays may miss the disk, especially at high $\theta_{\rm obs}$, when the disk appears more edge-on. As a result, we see more numerical noise at these angles. Nevertheless, with an increase in the resolution, errors are suppressed.

The high $\theta_{\rm obs}$ also produces a noticeable change in the spectrum continuum. In Fig.~\ref{fig:spec_spn_ang_rest}, we observe a rapid decrease in the continuum beyond the angle of $70^{\circ}$. The general relativistic light bending allows photons emitted from the far side of the disk, behind the black hole, to be redirected toward the observer. However, the flux observed from this lens region is significantly lower than the direct disk emission, as only a small fraction of photons are bent along paths that reach the observer at a given $\theta_{\rm obs}$, decreasing the overall continuum of the disk. Hence, the total line strength peaks around the region of 6.4 keV for lower $\theta_{\rm obs}$, approximately less than $20^{\circ}$, beyond which the contribution to the 6.4 keV region is not significant, and the peak either moves to higher energies or is completely smeared out.

\subsubsection{The influence of Lamp Height and Dissipation fraction}
\label{sec:height_energy_var}

\begin{figure}[!h]
    \centering
    \includegraphics[width=0.98\linewidth]{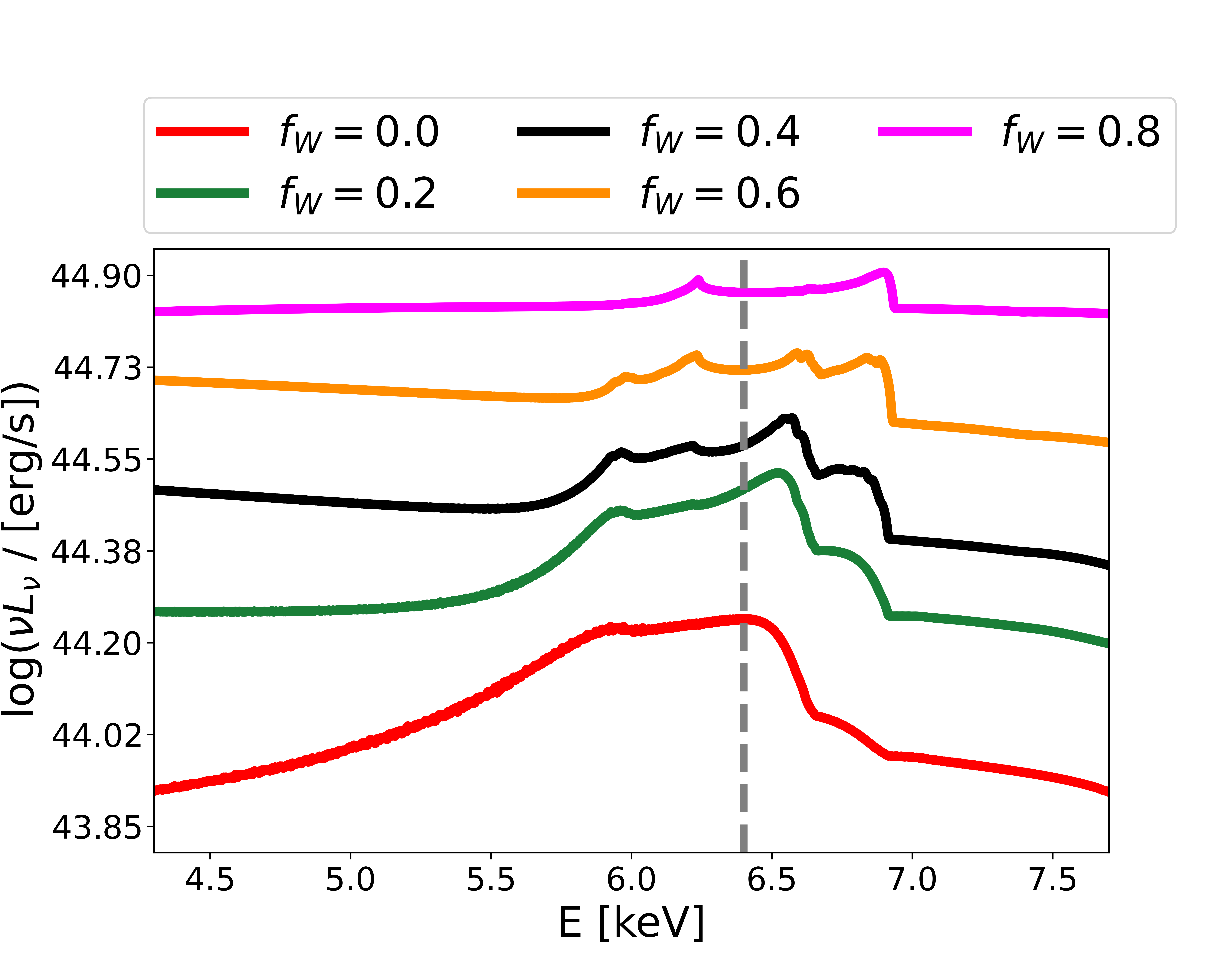}
    \caption{The variation of the Fe K$\alpha$ line profile with the dissipation fraction in the warm corona. The other parameters are set to their canonical values as shown in Tab.~\ref{tab:param_space}.}
    \label{fig:energy_var_hf}
\end{figure}

\begin{figure}[!h]
    \centering
    \includegraphics[width=0.98\linewidth]{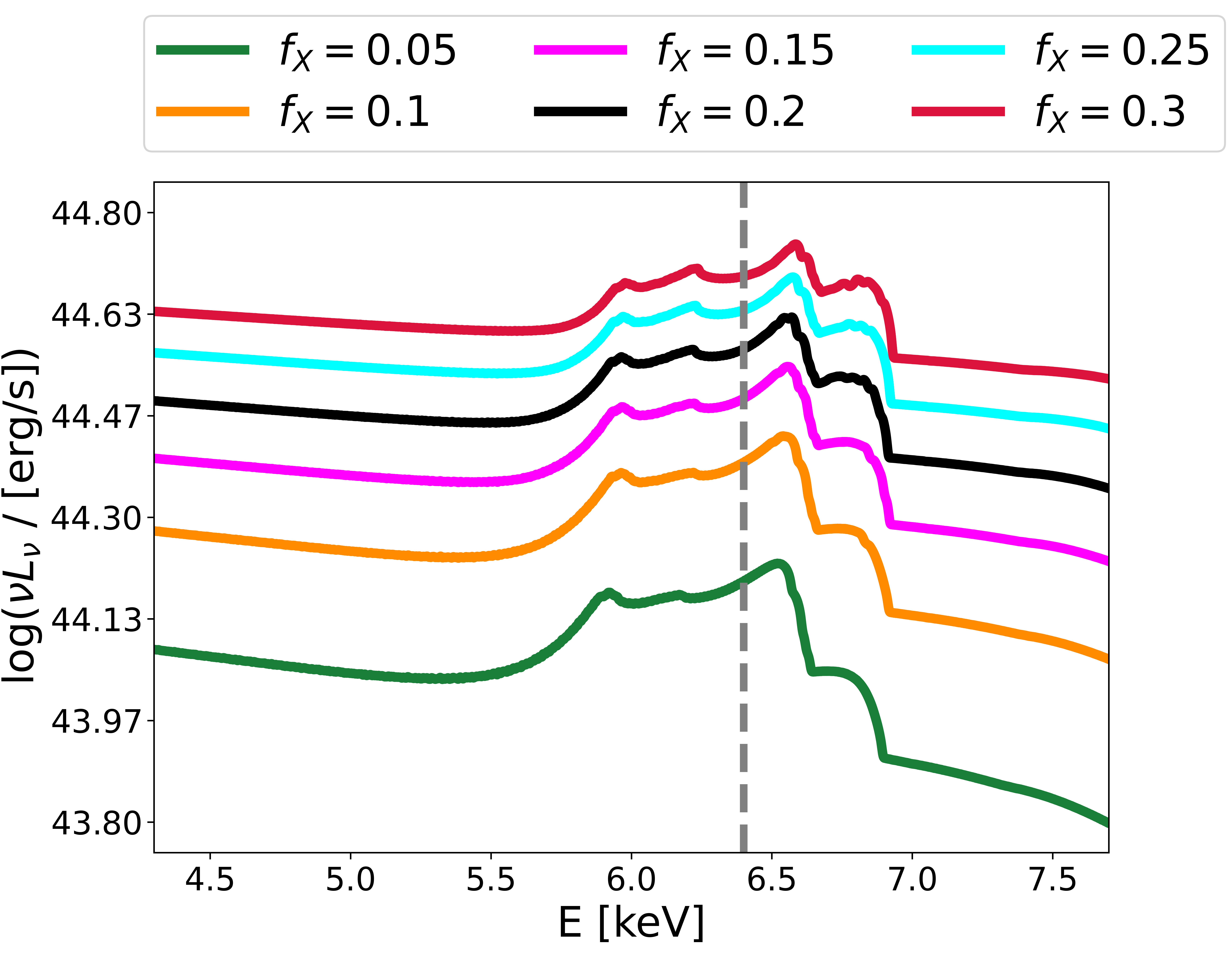}
    \caption{The variation of the Fe K$\alpha$ line profile with the illuminating X-ray flux fraction. The other parameters are set to their canonical values as shown in Tab.~\ref{tab:param_space}.}
    \label{fig:energy_var_fx}
\end{figure}

The position and energetics of the hot lamp and warm corona also determine the final line profile.
The top panel in Fig.~\ref{fig:height_var} shows how the relativistically corrected Fe K$\alpha$ line profiles respond to changes in the lamp height. Only slight differences are observed in the line shapes, indicating that $h$ has a relatively weak impact on the iron line shape. To understand the reason behind these small changes, we plotted the corresponding incident X-ray flux radial profiles $F_X(r)$ as a function of lamp height in the bottom panel of Fig.~\ref{fig:height_var}. While there are minor changes in the flux level, the inversion in the $F_X(r)$ appears around $10~{\rm r_g}$, consistent with the emissivity profile discussed by \citet{2012wilkins}. This behavior arises due to the underlying formalism, where the lamp height only affects the $F_X(r)$ function, which in turn changes $\xi(r)$, as defined in Eqs.~\ref{eq:FX_r} and~\ref{eq:xi_r}, respectively \citep{2017MNRAS.472L..60B}. 

We also see the variation of the energy flux distribution between the dissipative fraction ${ f_{W}}$ and incident X-ray Flux fraction $f_X$ in Fig.~\ref{fig:energy_var_hf} and Fig.~\ref{fig:energy_var_fx}, respectively. The variation of ${ f_{W}}$ gives the notion of the variation of the internal heating, which signifies the dissipative nature of the warm corona, while the variation of the $f_X$ represents the variation of the incident X-ray flux incoming from the compact hot corona above the SMBH. 

As we assume the energy flux $D(r)$ is varied between $f_X$ and ${ f_{W}}$, if ${ f_{W}}$ is varied, then the back illumination's contribution also changes, where $f_X$ is kept constant. Hence, as we change ${ f_{W}}$, we see 
in Fig.~\ref{fig:energy_var_hf} how the contribution of the internal heating affects the Fe~K$\alpha$ line profile. With no internal heating 
of the warm corona, i.e., ${ f_{W}} = 0$, thermal lines are still produced, as the temperature of the disk atmosphere is high enough to contain a large amount of highly ionised iron ions. With an increase of ${ f_{W}}$, we see the variation in the contribution from the FeXXV and FeXXVI. As ${ f_{W}}$ reaches higher values, the temperature of the matter structure rises, which ionizes FeXXV to FeXXVI, and the contributions from the FeXXVI increase. For an ${ f_{W}} = 0.8$, the FeXXV is almost fully ionized to FeXXVI, and hence we only observe the double-peaked contribution from the FeXXVI ions. Detailed contribution of different iron line transitions into the total observed line profile is discussed in the next section below.

A similar trend is also seen when we vary $f_X$, while ${ f_{W}}$ is kept constant. The FeXXV ions with an increase in $f_X$ ionize, and the strength of FeXXVI increases.  By increasing the emitting flux from the lamp, we observe an increase in the position of the continuum with the relativistically corrected Fe~K$\alpha$ profile. 
Hence, studying the variation in the highly ionized Fe K$\alpha$ profile in AGN sources can help us constrain the nature of the dissipative warm corona and the incident illumination strength. We also note that changes in the X-ray flux fraction $f_X$ or the lamp height $h$ over time can give insights into the reverberation studies of the accreting black hole systems. That is, the time-dependent changes in the reflected X-ray spectrum from the accretion disk, especially in the highly ionized Fe K$\alpha$ line, show how the disk responds to changes in the lamp above the SMBH.

\subsection{Contribution from different line transitions}
\label{subs:litr}

\begin{figure*}
    \centering
    \includegraphics[width=1.08\linewidth]{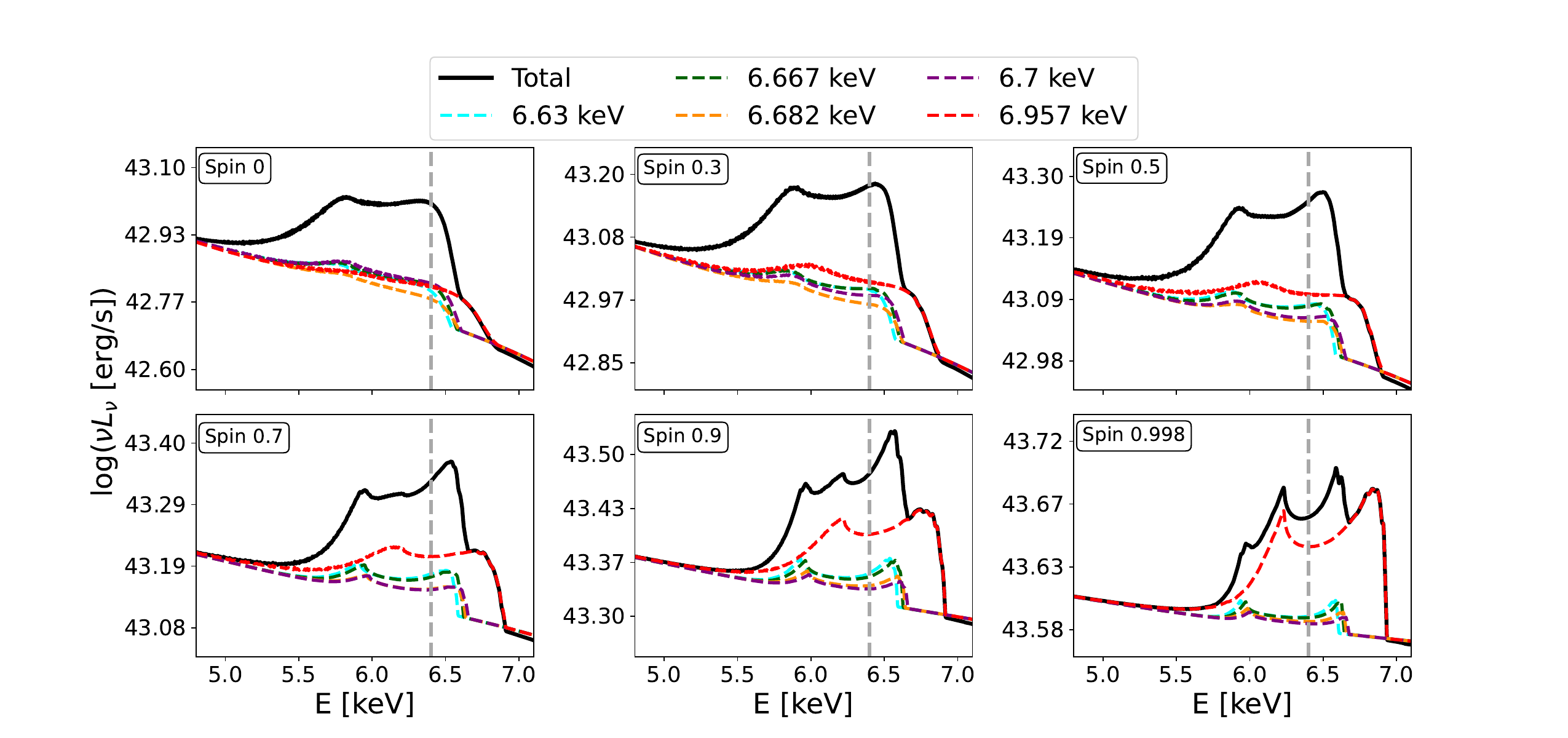}
    \caption{The contribution of major Fe K$\alpha$ lines to the total observed line profile is plotted by the black solid line. 
    Lines emitted locally at 6.63~keV (cyan), 6.667~keV (green), 6.682~keV (orange), and 6.7~keV (brown) are contributions from the different ionization states of FeXXV, while the line at 6.957~keV (red) is the contribution from the FeXXVI ion. The continuum level is arbitrarily lumped together for better visualization.
    Different panels are for various spin values marked in the boxes, all for $\theta_{\rm obs} =15^{\circ}$.  The vertical dashed line marks the position of 6.4~keV energy.}
    \label{fig:spec_ang_15}
\end{figure*}

\begin{figure*}
    \centering
    \includegraphics[width=1.08\linewidth]{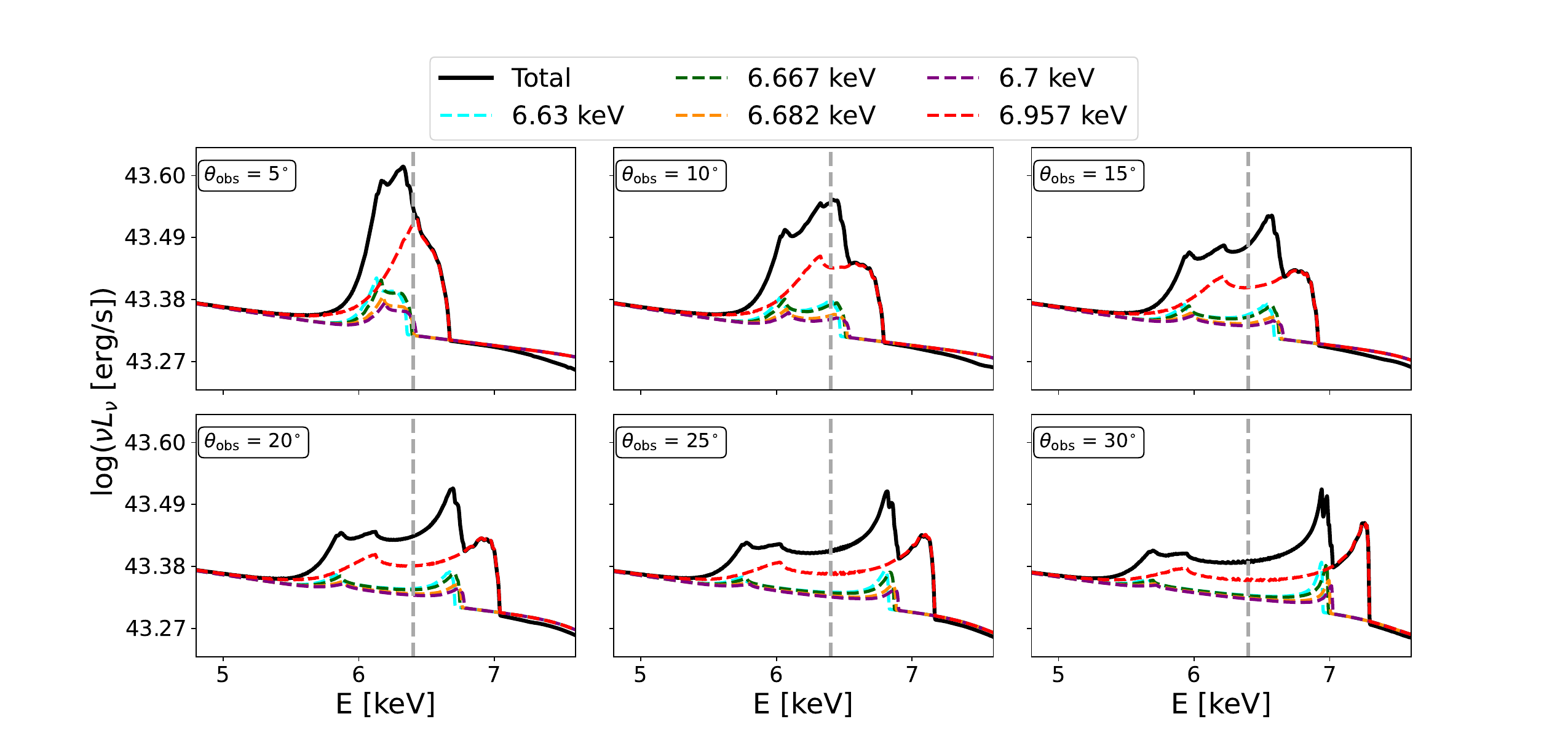}
    \caption{Same as in Fig.~\ref{fig:spec_ang_15}
    but for the single spin value of 0.9, and $\theta_{\rm obs}$ in the range from $5^{\circ}$ to $30^{\circ}$ given in the panel's boxes.}
    \label{fig:spec_spn_09}
\end{figure*}

\begin{figure*}
    \centering
    \includegraphics[width=1.08\linewidth]{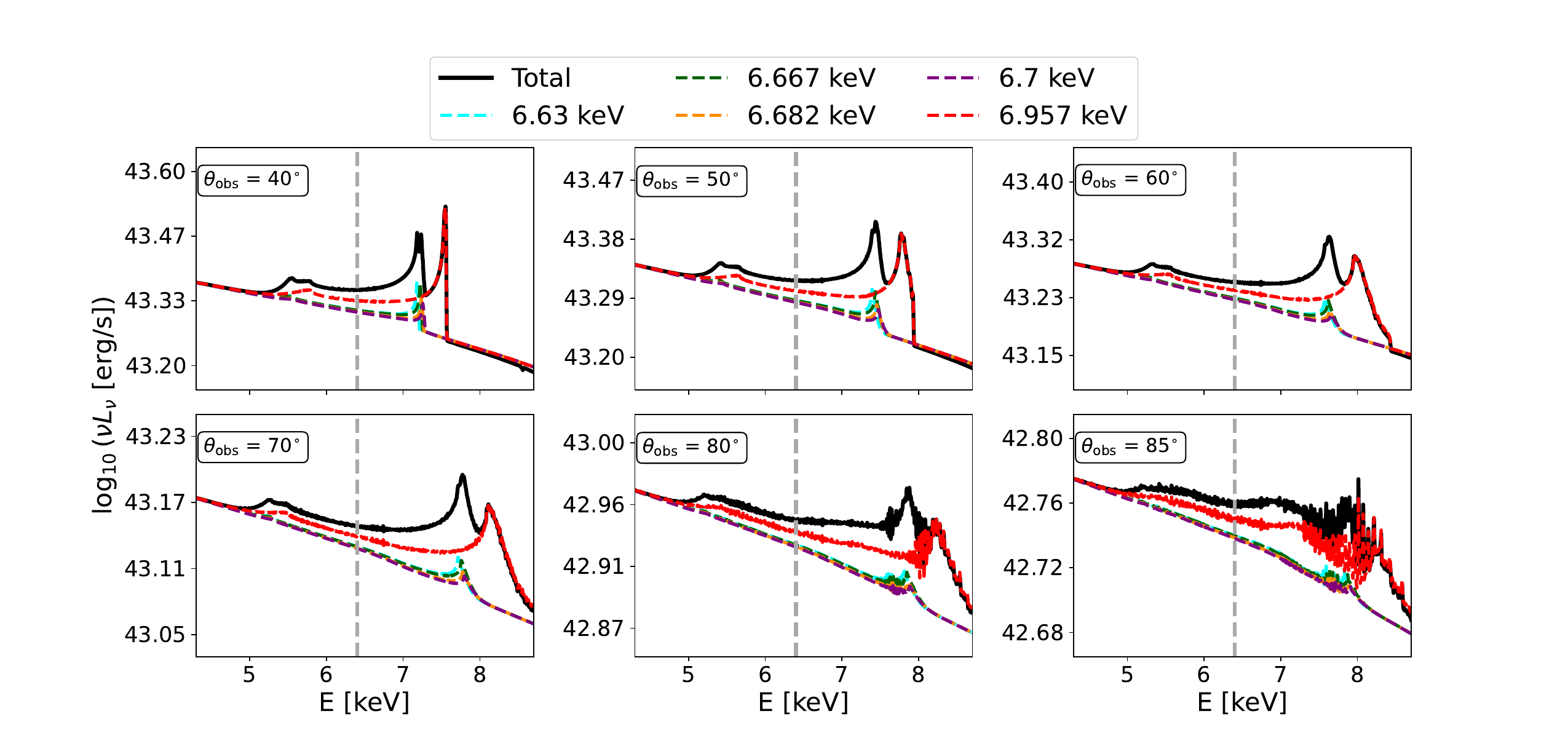}
    \caption{Same as in Fig.~\ref{fig:spec_ang_15}
    but for the single spin value of 0.9, and $\theta_{\rm obs}$ in the range from $40^{\circ}$ to $85^{\circ}$ given in the panel's boxes.}
    \label{fig:spec_spn_rest_09}
\end{figure*}

The contribution of the Fe K$\alpha$ lines that are emitted locally, to the total observed line profile is depicted in Fig.~\ref{fig:spec_ang_15} for a nearly face-on disk and six different spin values given in the boxes. 
With the increase in spin, the strength of the Fe K$\alpha$ line emitted due to FeXXVI at 6.957 keV increases, which is expected due to the rise in total temperature of the matter structure as shown in Fig.~\ref{fig:matter_structure}. 
The contribution of the FeXXV ions dominates most of the line profile, giving rise to the peak of the broad line structure at 6.4 keV for $\theta_{\rm obs} = 15^{\circ}$. Their shift is similar as these lines are generated very close to each other. A slight shift in their strength can be seen, suggesting the temperature is adequate for the matter structure to generate more FeXXV ions at the energies 6.63 keV and 6.667 keV, which correspond to the forbidden and intercombination lines, respectively \citep[see][]{2005MNRAS.357..599B}.  
For spin 0.998, we see the line structure has three distinctive peaks (bottom right panel of Fig.~\ref{fig:spec_ang_15}). Due to the high spin of the SMBH, the surrounding matter is heated to extremely high temperatures, leading to significant ionization. As a result, FeXXV ions become further ionized to FeXXVI, giving rise to a distinctive sharp peak for FeXXVI ions.

Figs.~\ref{fig:spec_spn_09} and~\ref{fig:spec_spn_rest_09} show the contribution of the highly ionized iron lines 
for rotating black hole with $a=0.9$ but across $\theta_{\rm obs}$ from $5^\circ$ to $30^\circ$, and from $40^\circ$ to 85$^\circ$, respectively.
The line profiles match the predictions in literature \citep[e.g.][]{2000PASP..112.1145F,2010MNRAS.409.1534D}. The effect of the tangential velocity of the disk is lower at $\theta_{\rm obs} = 5^{\circ}$ in comparison with $85^{\circ}$. Hence, with the increase in angle, the rotational Doppler effect starts to dominate, and the lines get broadened. The peaks corresponding to the lines from the highly ionized ions FeXXV and FeXXVI are very close to each other for lower $\theta_{\rm obs}$, and they are smeared into one peak. However, with increasing angle, the FeXXVI, which originates farther in energy than the FeXXV ions, starts showing a distinctive peak 
from $\theta_{\rm obs} \approx 40^{\circ}$. Beyond this angle, the Doppler effect is strong due to rotation, causing the lines to smear out. 

The disk outer radius for two-zone geometry is arbitrarily chosen in this paper. A natural question arises about fluorescent line emission from outer disk radii, where warm corona ends, and reflection from a cold disk may happen 
(see Fig.~\ref{fig:mod_params} lower right panel). This, so-called cold iron line, may contribute 
as an additional component to the total broad line profile seen by the distant observer. In Appendix~\ref{Appx:cold_line}, we clearly show that due to the radial emissivity profile interpolated from \texttt{TITAN} calculations, the cold iron line has a negligible contribution to the total line profile.
For the complete explanation of the broad iron line profile, we wish to physically define the transition radius between two-zone (warm corona/cold disk) and 
one-zone (cold disk) accretion flows. This issue will be the goal of our future research.

\section{Discussion and Conclusion} \label{sec:dic_con}
In this paper, we demonstrated that the broad feature usually observed in AGN around the 6.4~keV also has contributions from the highly ionized Fe K$\alpha$ lines. This can be modeled by ray-traced emission from a two-slab system containing a warm corona on the top of a cold disk, which is externally illuminated by hard X-ray power-law radiation from a lamp located above the central black hole.
The inner region of the disk, surrounded by a warm corona, is fully dissipative and 
reaches very high temperatures at large optical depth, in line with the recent theoretical and observational predictions \citep[e.g.][]{2022MNRAS.515..353X,2023gronkiewicz,2024A&A...690A.308P}. Under such high temperatures, iron is highly ionized \citep{2020MNRAS.491.3553B}, and the emission lines corresponding to these ions, so-called iron line complex, are created at higher local energies than 6.4 keV.
Nevertheless, due to the close vicinity of the SMBH, these lines broaden and become gravitationally redshifted. Depending on the observer's viewing angle, the highly { ionized} Fe K$\alpha$ lines produced in the warm atmosphere may shift near the 6.4~keV, 
and all of them partially participate in the total broad line profile. 
{ For the purpose of this paper{ ,} we decided to concentrate only on line broadening due to relativistic effects, while Compton broadening will be included in the next step of our studies.}

We considered two-dimensional matter stratification to investigate the relativistic effects on highly ionized Fe K$\alpha$ iron lines generated from FeXXV and FeXXVI ion populations. 
The radial solution adopted was taken from 
\citet{2022MNRAS.515..353X} and \citet{2024MNRAS.530.1603B}, where it was used to formulate the \texttt{reXcor} model to reproduce the soft X-ray excess widely observed in AGN. In the \texttt{reXCcor} model, local reflected spectra have been computed using a photoionization code designed by \citet{1993Ross} and later modified by \citet{2001Ballantyne}. We follow this approach; nevertheless, for the solution of the matter structure and radiation transfer, we used the  
\texttt{TITAN} code by \citet{2003A&A...407...13D}. 
It was demonstrated, that \texttt{TITAN} is appropriate for matter under broad ionization conditions { and being additionally heated by internal gain of energy, as it is required in the warm corona} \citep{2020A&A...634A..85P, 2024A&A...690A.308P}. 

Our study shows the case of AGN with SMBH of the assumed mass $10^8\ {\rm M_{\odot}}$, and with a disk of the mass accretion rate equal to 0.1 in units of Eddington accretion rate. According to the radial model used, the total energy flux generated due to accretion is fractionally distributed among three regions: hot lamp, warm dissipative corona, and cold disk. We show the dependence of the highly ionized Fe K$\alpha$ features on the intrinsic parameters: spin, viewing angle, lamp height, dissipative fraction, and the illuminating X-ray flux fraction.

The central SMBH spin plays a significant role in shaping the matter structure and influencing the relativistic shifts of emission lines, since we assume that the inner edge of the accretion disk is located at the $r_{\rm ISCO}$, given solely by the spin. As the spin increases, the line profile becomes more pronounced, often exhibiting a more distinctive peak structure. While this may seem counterintuitive, the primary cause of these changes is not the variation in disk rotation, but rather the alteration in the disk's vertical temperature structure, thereby changing the strength of the local emission from the highly ionized Fe K$\alpha$. The SMBH's spin has minimal impact on disk rotation at large radial distances, but it becomes crucial in the innermost 
regions near the SMBH, by increasing the emitting surface with relatively high temperature that ensures a large FeXXV and FeXXVI ion population. This is a crucial factor as the temperature is heavily spin-dependent. Hence, for higher temperatures, the contribution from the FeXXVI ion increases, giving rise to a distinctive peaky structure in the line spectrum.



 As expected, the viewing angle ($\theta_{\rm obs}$) of the observer plays an important role in the overall observed spectral shape.
 For a lower observing angle, the line shift is prominent at 6.4~keV, but with an increase in angle, the broad lines extend beyond 6.4~keV. At high $\theta_{\rm obs}$ (above $\sim 30^\circ$), the lines are affected by extreme relativistic effects. Hence, for low $\theta_{\rm obs}$, the broad lines observed near 6.4 keV \citep[see][]{2006AN....327.1032G} in AGN sources may contribute to the signatures of broadened and relativistically redshifted iron lines emitted from the FeXXV and FeXXVI ion populations. 
 
 We also investigated the lamp height and the distribution of the total energy flux. We reported that the lamp height did not show a significant change in the iron line profile. Hence, we studied how the illuminating X-ray flux changes with the height of the lamp. We noticed that the flux tends to pivot around a radial distance of about $10$–$20\,{\rm r_g}$. This means that when the lamp is close to the SMBH, the inner regions of the disk receive more X-ray light than the outer parts. This happens due to strong gravitational effects that bend the light more toward the center, as shown in Eq.~\ref{eq:FX_r}. However, when the lamp is placed high above the black hole, the effect of gravity is weaker. As a result, the flux spreads out more evenly and decreases smoothly with distance from the center. 
 
 On the other hand, the changes in the fractions of the energy flux distribution among the three regions have a significant effect on the iron line profiles. The increase of internal heating in the warm corona first results in a higher contribution from FeXXV, and further heating ionizes the FeXXV to FeXXVI. This shows that effects such as changes in viscosity or magnetic field strength that induce changes in the dissipation of the warm atmosphere can highly affect the iron line profile. We also observe a strong correlation between the X-ray illumination and the iron line profile. An increase in the X-ray flux emitted from the lamp alters the profile of the iron line, particularly affecting the relative contributions of FeXXV and FeXXVI. This reflects how variations in the internal properties of the lamp can induce changes in the accretion disk that resemble reverberation effects. Although we focus here solely on the broad Fe K$\alpha$ feature, and a direct analysis of reverberation is beyond the scope of this study, the observed changes in the line profile may still serve as a useful diagnostic of the lamp's influence on the X-ray spectrum.

Numerous studies have shown the presence of { the} broad iron line structure { in X-rays} for different AGN sources \citep[e.g.][ and references therein]{2000PASP..112.1145F,2006AN....327.1032G,2007ARA&A..45..441M}. MGC-6-30-15 is one of the best candidates that shows a broad emission line in its X-ray spectrum \citep{1995Natur.375..659T}. On a qualitative comparison, we see that our results for viewing angles of $20^{\circ} - 30^{\circ}$ match the MGC-6-30-15 dataset shown in Fig. 4 of \citet{2007ARA&A..45..441M}. In \citet{1996MNRAS.282.1038I}, a disk line is fitted assuming an angle of $30^{\circ}$ obtained from \citet{1995Natur.375..659T}. These results show a strong resemblance to the scenario of highly ionized iron ions getting redshifted and contributing near the 6.4 keV region.

We note that in our work, we have included the relativistic effects, but as a limitation of \texttt{TITAN}, we have not included the effects of Comptonization on the microscopic level. We are aware that multiple Compton scatterings of the line originating from dense matter can modify the final line profile; nevertheless, for this paper, we aimed to clearly show only one phenomenon that changes the line profile, which is the strong gravity. Adding proper treatment of Compton scattering will be included in our future work. The next important limitation of our model is the position of an outer cold disk from which the reflection may also occur and contribute to the overall spectrum. In the current model, we do not consider the reflection from the cold outer disk, since according to our studies see Appendix~\ref{Appx:cold_line}, the final line profile will be fully dominated by the emission from the inner disk's warm corona region. We note that the above conclusion depends on the transition radius between a two-zone disk plus corona accretion flow and a one-zone cold disk. And only one, arbitrarily assumed, radial transition was considered in this paper. In our future work, we plan to develop a physically consistent model where the warm corona can build up on the top of a cold disk, and the above transition will be self-consistently determined. This paper presents a complete method that can be used as soon as we can change the transition radius. Moreover, the reverberation studies will be possible in such a complete model.

{ Contrary to assumption of the soft excess as an ionized standard disk emission \citep{2010garcia, 2011garcia, 2014garcia}, we assume the disk to have an additional heating component. We adopt the radial profile of internal heating from \citet{2022MNRAS.515..353X}, while noting that the physical origin of the dissipative nature of warm corona is possibly linked to magnetic fluxes which have been discussed by \citet{2023gronkiewicz}. The deviation of the radial dependence of the emissivity profile from the standard disk assumption (as depicted in Appendix \ref{Appx:cold_line}) may also show the existence of the warm corona to have a dissipative nature. We also show the importance of the inclusion of the relativistic effects from the emission of the warm corona, which is also discussed by \citet{2014garcia}. Hence, the broad line component observed in the X-ray spectrum can be described via the emission from the warm corona.}

We also aim to study further the effect of the mass, accretion rate, viscosity parameter, the radiative efficiency, and finally, the matter structure on the emitted iron line profile.
As future work, we aim to build a larger parameter space with larger constraints to compare it with the upcoming data from observatories like XRISM and 
NewATHENA. Broad iron line profiles have already been observed in multiple AGN sources \citep[e.g.][]{2001A&A...365L.134R,2001ApJ...559..181P,2006AN....327.1032G}. We aim to develop a model that offers a better parameterization for constraining the spin, mass, and viewing angles of SMBH in AGN.

\begin{acknowledgements}
     PPB has been fully and AR has been partially supported by the Polish National Science Center grant No. 2021/41/B/ST9/04110. All computations have been performed using our computer cluster at NCAC PAS. DL acknowledges the Czech Science Foundation (GA\v{C}R) grant no. 25-16928O. We would like to thank David Ballantyne, Xin Xiang (Cindy) 
     and Biswaraj Palit for their insightful comments on the paper.
\end{acknowledgements}

\bibliography{sample631}{}
\bibliographystyle{aa}

\begin{appendix} 

\section{Single Line Test} 
\label{Appx:single_lin}

\begin{figure}[!h]
    \centering
    \includegraphics[width=1.0\linewidth]{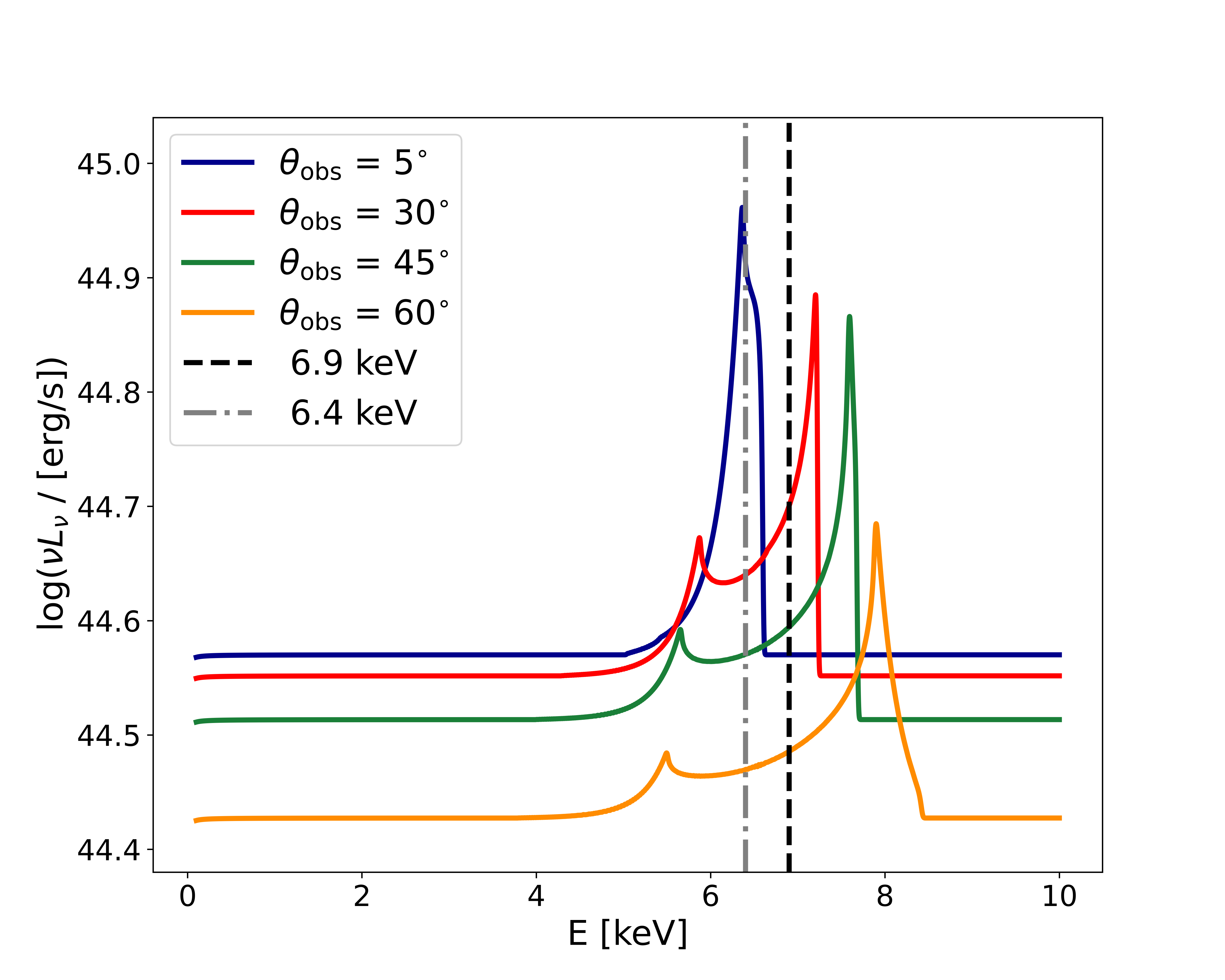}
    \caption{Spectrum generated by \texttt{GYOTO} for various viewing angles, computed for an injected single Gaussian line with $\mu = 6.9\, {\rm keV}$ and $\sigma=10\, {\rm eV}$. The computations have been made for a non-spinning black hole with $10^8\, {\rm M_{\odot}}$, and for integration disk surface from $r_{\rm in} = r_{\rm ISCO} + 1$ to $r_{\rm out} = 25\, { r_{\rm g}}$. The equivalent width was chosen as $600\, {\rm eV}$ and was kept constant at all radii. The vertical dashed black line depicts the 6.9 keV, while the vertical dash-dotted gray line depicts 6.4 keV.}
    \label{fig:slt_spec}
\end{figure}
Here we present the results for the single line test, where we inject a 6.9 keV Gaussian function with the flat continuum at all the disk radii. This scenario is unrealistic as the line flux is kept constant throughout the disk radial structure with an arbitrary value of the equivalent width of 600 eV. We aimed to show that the inclusion of relativistic effects on the highly ionized Fe K$\alpha$ line may contribute to the broad line profile observed at the energy of the neutral (6.4 keV) line.

 Fig.~\ref{fig:slt_spec} shows the relativistically corrected spectrum for the case of a non-spinning SMBH, varying with the viewing angle $\theta_{\rm obs}$. We observe the profiles to follow a similar structure as described in \citet{2010MNRAS.409.1534D} and \citet{2024arXiv241114338G}. For a low viewing angle, the shift of the line from the 6.9 keV (marked with a dashed vertical line in Fig.~\ref{fig:slt_spec}) to the energy of the neutral fluorescent line (dash-dotted vertical line in Fig.~\ref{fig:slt_spec}) is observed. Additionally, with a low $\theta_{\rm obs}$, we see a high continuum flux as more photons reach us from a face-on view, which is also consistent with the previous results \citep[e.g.][]{2000PASP..112.1145F}. The line shape broadens with an increase in $\theta_{\rm obs}$ as the tangential component of the rotational velocity contributes more, which leads to extreme line broadening.  Single line test shows the correctness of the \texttt{GYOTO} code, but it does not include a proper radial emissivity profile of a line flux. It corresponds to the disk line models of \citet{1989fabian} and of \citet{1991laor} that are available in the commonly used XSPEC fitting package \citep{1996arnaud}. To study the realistic emissivity profile of the iron line complex together with the underlying continuum, we used a combination of \texttt{TITAN} and \texttt{GYOTO}, with the results described in Sec.~\ref{sec:results}.

\section{Reflection from an external cold disk} 
\label{Appx:cold_line}

\begin{figure}[!h]
    \centering
    \includegraphics[width=1.08\linewidth]{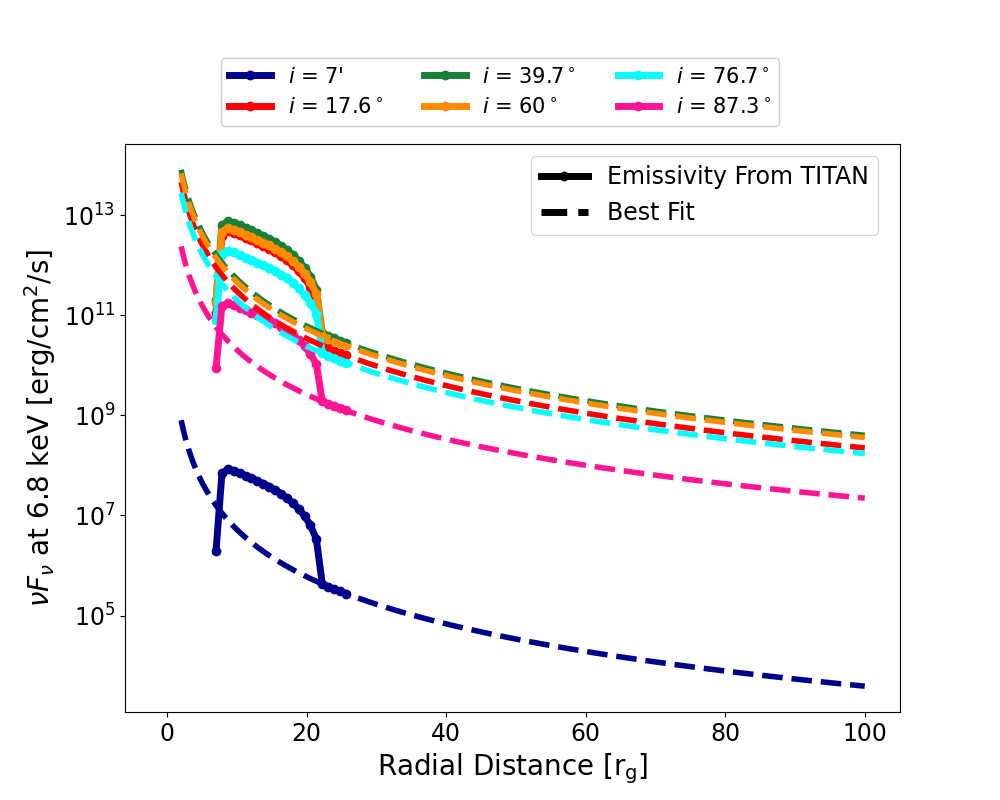}
    \caption{The variation of the flux emitted at 6.8 keV with the radial distance. The computations were done for different emission angles ($i$) prescribed in \texttt{TITAN}.}
    \label{fig:emmi}
\end{figure}

The two-zone geometry, {  which} includes the warm, optically thick corona above the cold disk, ends on an arbitrarily chosen outer radius 25$\, r_{\rm g}$. Since we do not know the physical mechanism that causes the warm corona disappearance, we decided not to explore this parameter in the current paper. Our future plan is to explore this issue toward { a} complete solution of the disk/corona geometry. Nevertheless, we should not completely forget about the reflection from the outer cold disk and eventual production of the fluorescent iron line.
Hence, to test here the contribution from the outer region of the disk, we appended additional disk rings up to $100\, r_{\rm g}$. The local emission from those rings contains a continuum with a Gaussian-like shape of emission line originating from a fluorescent process. We tested the contribution for the outer region only in the case of a non-rotating SMBH, i.e., $a=0$. 

\begin{figure}[!h]
    \centering
    \includegraphics[width=1.08\linewidth]{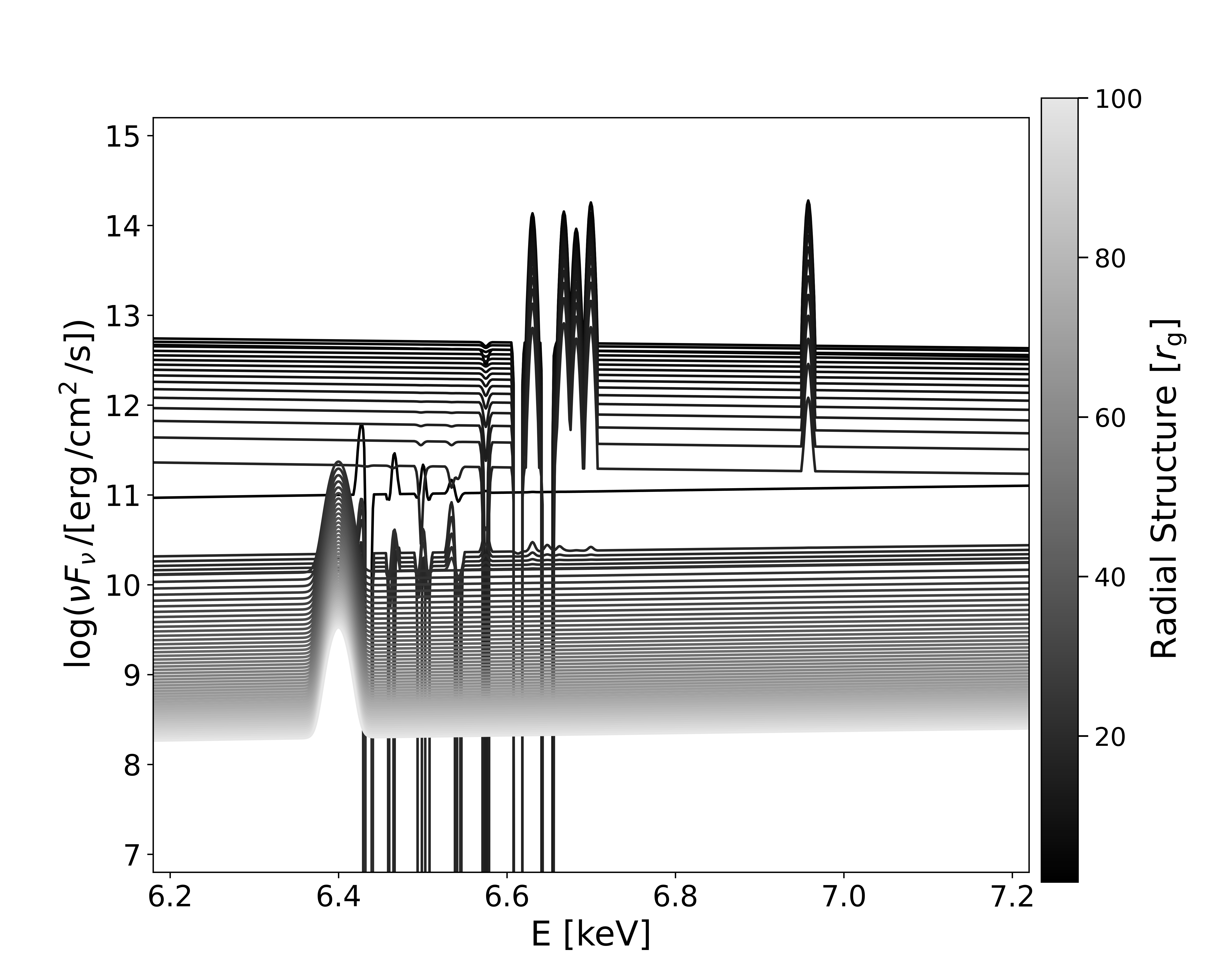}
    \caption{The output spectrum from \texttt{TITAN} added with the  spectra of the best fit parameters beyond 25$\, r_{\rm g}$. The 6.4 keV line is injected with a 400 eV equivalent width. This figure is depicted for the emission angle $i = 17.6^\circ$.}
    \label{fig:titan_6.4kev}
\end{figure}
 \begin{figure*}[!h]
    \centering
    \includegraphics[width=1.08\linewidth]{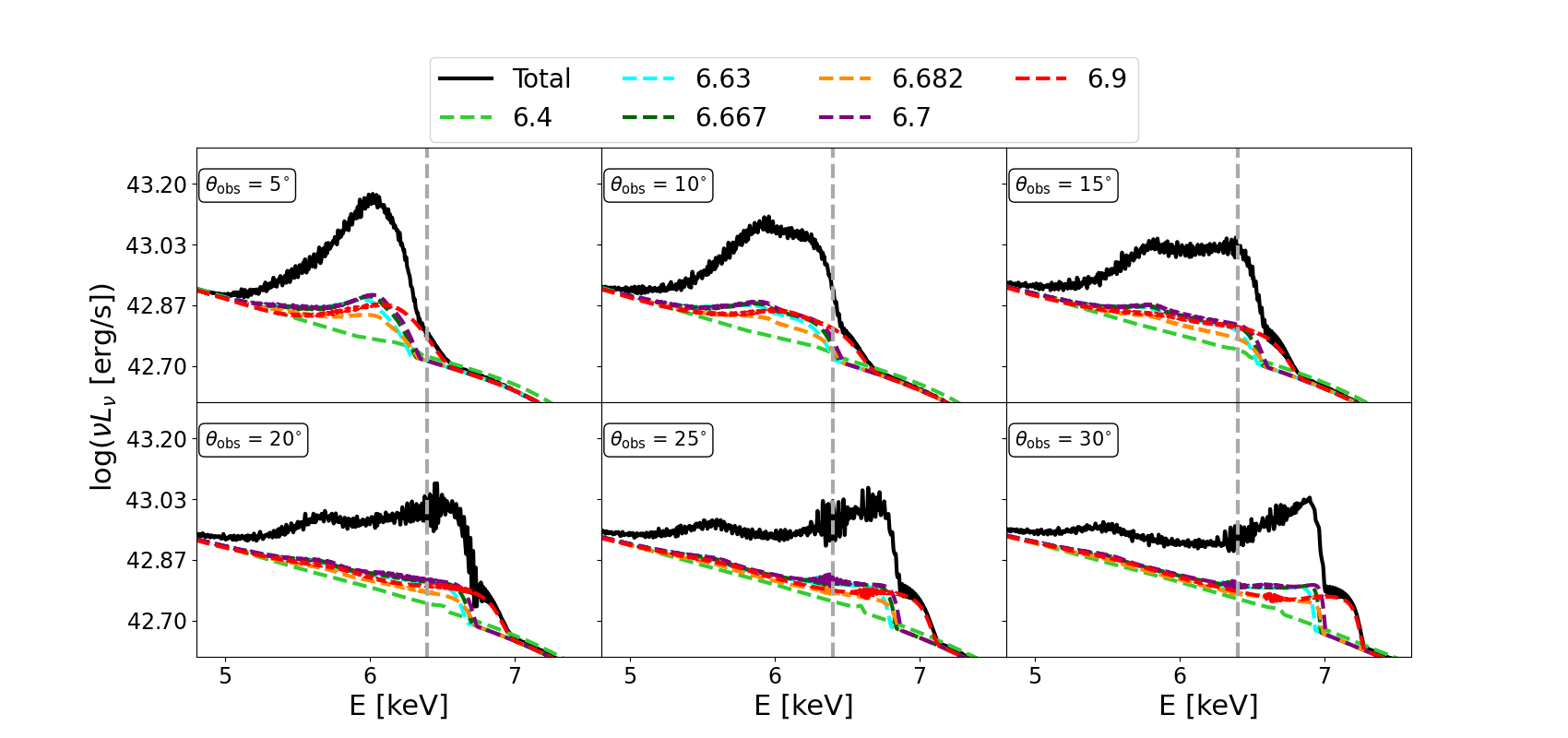}
    \caption{The contribution of the individual lines to the total line profile computed for different angles. The depicted colours are the same as Fig.~\ref{fig:spec_ang_15} with the addition of light green showing the contribution from the 6.4 keV fluorescent iron line.}
    \label{fig:signl_contri_appxb}
\end{figure*}

 As a starting point, we have to estimate the radial emission profile (EP) up to the outer radius $100\, r_{\rm g}$. To extract EPs, we take radial flux values of reflected spectra calculated by \texttt{TITAN} code for different emission angles and extrapolate them to larger radii by fitting the following emissivity law:

 \begin{equation}
     {\rm EP}= K (r/r_{\rm g})^{n}, 
     \label{equ:eprofile}
 \end{equation} 
 where $K$ is a power law normalization constant, and $n$ is an emissivity power law index. We plot the EPs resulting from   \texttt{TITAN} code by a solid line in  Fig.~\ref{fig:emmi}, where we choose continuum flux 
 at 6.8~keV as a reference point. The region where the warm corona is dominated by Compton cooling is seen by the strong 
 radial bump in the emissivity. This may be a potential radial size of a warm corona, but we plan to explore this issue in our future research. The bump decreases with radius, and the EP given by Eq.~\ref{equ:eprofile} is fitted to the last 5 points, which depict the region where line cooling starts to dominate, as discussed in Sec.~\ref{res:ms}. 
 For all the profiles, we found that the best-fit for emissivity follows a profile close to $r^{-3}$, which depicts the EP for a standard  \citep{2012wilkins}. The best fit values for each emission angle are shown in Tab.~\ref{tab:params_emmi}.

\begin{table}[]
    \centering
    \caption{Best fit parameters for the angle-dependent Emissivity Profile given by Eq.~\ref{equ:eprofile}, and depicted in Fig.~\ref{fig:emmi}. Here $i$ is the emission angle from \texttt{TITAN}. The best-fit parameter is computed for the continuum emission at 6.8 keV.}
    \begin{tabular}{ccc}
    \hline
    
         $i$ & Normalization $K$   & Emissivity Power Law $n$\\
         & &              \\
         \hline
         \hline
          & & \\
        7'              &   6.98e+09    &   -3.1276 \\
        $17.6^{\circ}$  &   4.02e+14    &   -3.1273 \\
        $39.7^{\circ}$  &   7.01e+14    &   -3.1224 \\
        $60^{\circ}$    &   5.88e+14    &   -3.1058 \\
        $76.7^{\circ}$  &   2.35e+14    &   -3.0679 \\
        $87.3^{\circ}$  &   1.85e+13    &   -2.9612 \\
        \hline
    \end{tabular}
    \label{tab:params_emmi}
\end{table}

Further, we construct the continuum for the radial range 25-100$\, r_{\rm g}$ with 50 points in between using the reflected continuum at 25$\, r_{\rm g}$ to follow the best fit power law of EP. Then we add a 400 eV line at each within this radial distance. In this way, we created an expected level of continuum and line spectra for the outer cold disk region (i.e., without a warm corona). The dependence of continuum levels and lines on radii for the case of emission angle 
$i=17.6^{\circ}$ is presented in Fig~\ref{fig:titan_6.4kev}. The fluorescence iron line at 6.4~keV is visible as a reflection from outer radii.

 The final ray-traced spectrum is computed after integration of the emission up to  $100\, r_{\rm g}$. The contribution of each iron transition to the total broad observed feature is presented in  
  Fig.~\ref{fig:signl_contri_appxb}, similar to Sec.~\ref{subs:litr}. 
  Cold iron line (depicted by green dashed line) has a negligible impact on the final broad iron line profile. 
  Hence, the two-zone (warm corona on the top of a cold disk) region within 25$\, r_{\rm g}$ is extremely important, which contributes to the formation of the broad iron line spectrum. 
  More cases will be considered in our future papers, when  
  the radial extension of the warm corona will be self-consistently defined. 

\end{appendix}

\end{document}